\pdfoutput=1
\documentclass[a4paper,onecolumn,unpublished,nopdfoutputerror]{quantumarticle}
\usepackage{amsmath,amssymb,amsthm,mathtools}
\usepackage{braket}
\usepackage{booktabs}
\usepackage{array}
\newcolumntype{L}[1]{>{\raggedright\arraybackslash}p{#1}}
\usepackage{url}
\usepackage{xcolor}
\usepackage{float}
\usepackage{algorithm}
\usepackage{algpseudocode}
\usepackage{tikz}
\usetikzlibrary{positioning, arrows.meta, fit, backgrounds, calc}
\usepackage{standalone}
\newcommand{\rtag}[1]{{\scriptsize\itshape\mdseries #1}}
\usepackage[font=small]{caption}   %
\usepackage{hyperref}
\hypersetup{hidelinks,pdftitle={Exact sampling beyond Fock inputs and linear optics},pdfauthor={Jean Senellart, Stephen Wein, Pierre-Emmanuel Emeriau}}
\DeclareMathOperator{\per}{Per}
\DeclareMathOperator{\Tr}{Tr}
\newcommand{\C}{\mathbb C}
\newcommand{\E}{\mathbb E}

\newcommand{\perrect}{\per_{\square}}

\newtheorem{lemma}{Lemma}
\newtheorem{proposition}{Proposition}

\theoremstyle{definition}
\newtheorem{definition}{Definition}
\theoremstyle{remark}
\newtheorem{remark}{Remark}

\title{Exact sampling beyond Fock inputs and linear optics: photon-number superpositions, photonic
matrix-product states, adaptive and non-linear circuits}
\author{Jean Senellart}
\affiliation{Quandela, 7 Rue L\'eonard de Vinci, 91300 Massy, France}
\author{Stephen Wein}
\affiliation{Quandela, 7 Rue L\'eonard de Vinci, 91300 Massy, France}
\author{Pierre-Emmanuel Emeriau}
\affiliation{Quandela, 7 Rue L\'eonard de Vinci, 91300 Massy, France}

\begin{document}
\sloppy
\maketitle

\begin{abstract}
Clifford and Clifford's algorithm samples single-photon boson sampling exactly, one output mode at a time, by adding
the input photons in a uniformly random order. We extend this construction to other input states and to adaptive and non-linear circuits. It applies whenever
non-negative weights on output prefixes and input sets satisfy one marginal identity, a set-chain lemma: the
random-order sampler is then exact at one weight evaluation per candidate mode, and it gives the posterior over the
hidden input set. We build four exact
samplers on it.
For sources in a superposition of vacuum and one photon, the weights are rectangular permanents on an offset
input set, and the cost has the exponent of single photons under loss. For photonic matrix-product states of bond dimension
$\chi$, a gap-by-gap Kraus contraction turns each weight into a short sum of squared amplitudes. For adaptive linear optics, the
chain starts from the posterior given the adaptive counts and rejects chains that contradict them. For a single-mode non-linear
phase the identity fails, and rejection from linear branches is exact with acceptance $1/\|c\|_1^2$. All four samplers agree
with strong simulation on small instances, and cost models put adaptive sampling with up to three adaptive modes within $10\%$
of single photons and a non-linear site at $\varphi=\pi/2$ at about $4\times$.
\end{abstract}

\section{Introduction}
\label{sec:intro}

Claims of quantum advantage with photons rest on the gap between what a device samples and what a classical computer can
sample exactly. For standard boson sampling, $n$ single photons in an $m$-mode linear interferometer $U$, the output
probabilities are permanents, \#P-hard to compute in the worst case, and a polynomial-time exact classical sampler would
collapse the polynomial hierarchy~\cite{AA2013}. Real photonic devices differ from this model: their sources can be partially
excited or entangled, and their circuits can include mid-circuit measurements with feed-forward or non-linear elements. Where
exact classical simulation stops for such devices depends on exact samplers built for them.

Clifford and Clifford~\cite{CC2018} gave the standard exact sampler for single photons. They observed that if the photons are
given a uniformly random order, the marginal law of the output modes of the first $j$ photons (the first $j$ output
\emph{rows}) is the average, over the $j$ input columns those photons used, of a single squared permanent. The output can then
be drawn row by row, and a sample costs about as much as two permanents of the full size, $O(n2^n)$. Their 2020
refinement~\cite{CC2020} exploits repeated output rows and brings the average cost over Haar interferometers to
$O(n\varrho^n)$, the growth $\varrho$ depending on the mode ratio $m/n$ ($\varrho\approx1.81$ at $m=2n$).

The proof uses three properties of this setting: the input is a product of single photons; the interferometer acts on them as
$U^{\otimes n}$; and the columns of $U$ are orthonormal, $\sum_rU_{rx}\bar U_{rx'}=\delta_{xx'}$, the identity that turns the
Laplace expansion of a squared permanent along its last row into a sum over the permanents of the smaller column sets. We
consider four kinds of devices. The first two break the product input, the last two the action $U^{\otimes n}$:
\begin{itemize}
\item \emph{Superposed sources.} Partially excited sources emit a coherent superposition of vacuum and one
photon~\cite{Renema2020SBS}, so the photon number is not fixed.
\item \emph{Photonic matrix-product states.} A quantum emitter with a $\chi$-dimensional internal memory, such as a quantum-dot
spin, emits photons one after another, entangled through that memory: a photonic matrix-product state (MPS) of bond dimension
$\chi$~\cite{Schon2005}, such as linear cluster and caterpillar states~\cite{Lindner2009,Huet2025}.
\item \emph{Adaptive linear optics.} Mid-circuit photon counting with feed-forward makes the interferometer depend on earlier
outcomes~\cite{CMS2021,OhLim2025adaptive}.
\item \emph{Non-linear sites.} A single-mode non-linearity between interferometers makes the circuit itself
non-linear~\cite{Spagnolo2021}.
\end{itemize}
None of them breaks orthonormality directly, but it is the property that fails inside the last two samplers: on the
non-adaptive modes of an adaptive circuit, and between the linear branches of a non-linear one
(Secs.~\ref{sec:adaptive} and~\ref{sec:nlbs}). Uniform loss and partial distinguishability in the mixture model are handled by every sampler, as a
mixture drawn before the chain (Sec.~\ref{sec:lemma}).

\paragraph{Prior work.} Renema~\cite{Renema2021marginals} writes the chain rule for an arbitrary input state. Its marginal is
a double sum over input components and hidden subsets, so a step costs up to $\binom nk^2$ component pairs for $n$ superposed
sources with $k$ photons, and a number exponential in the length of the train for an emitter. Dias and
K\"onig~\cite{DiasKoenig2025} sample any measurement of a coherent superposition by rejection from the mixture of its
components, with an expected number of proposals set by the number of components ($2^n$ for $n$ superposed sources) and the
2-norm of their coefficients. Rejection or reweighting over a decomposition into classically easy terms, at a cost set by the
1-norm of the coefficients, is also the principle of quasi-probability and stabilizer-rank simulators of quantum
circuits~\cite{HowardCampbell2017,Bravyi2019,Seddon2021}. Oh and Lim~\cite{OhLim2025adaptive} sample adaptive circuits exactly
and efficiently when the input and the final measurement are Gaussian and $O(\log m)$ modes are read adaptively. Bravyi, Gosset
and Liu~\cite{BGL2022} sample any circuit from prefix amplitudes, without marginals (cost models compared in the Supplementary
Material, Sec.~\ref{S-sec:compete}). Moylett et al.~\cite{Moylett2020} build on the Clifford--Clifford sampler to simulate
single photons under loss and partial distinguishability. Further exact methods --- repeated input modes~\cite{BrodOszmaniec},
Gaussian chains~\cite{QuesadaArrazola2020,Quesada2022,Bulmer2022}, bounded treewidth~\cite{OhTreewidth2022}, phase-space
representations~\cite{RahimiKeshari2016} and a linear-optical Kerr gadget~\cite{Spagnolo2021} --- are compared in the
Supplementary Material (Table~\ref{S-tab:prior}). To our knowledge none of them specialises the random-order chain to
superposed sources, to photonic MPS with a cost set by the bond dimension and the number of traced runs, or to Fock inputs with
photon-number-resolved feed-forward and final detection.

\paragraph{Contributions.} We state the property the random order provides as a lemma about set-indexed weights, which also
gives the posterior over the hidden input set (Sec.~\ref{sec:lemma}), and build four samplers on it:
\begin{enumerate}
\item for superposed sources, a \emph{rectangular permanent} on an active set started at $n-k$ sources, at the cost exponent of
single photons under loss $\eta=\sin^2\alpha$ (Sec.~\ref{sec:sbs}, Proposition~\ref{prop:sbs});
\item for photonic MPS, a \emph{gapwise Kraus contraction} that writes each weight as $R_j$ squared amplitudes evaluated by
matrix-valued Glynn sums; on the trains we measured, $R_j$ is $1$--$2$ on the last steps, which carry the cost
(Sec.~\ref{sec:ebs}, Proposition~\ref{prop:gap});
\item for adaptive linear optics, a \emph{conditional chain} started from the lemma's posterior given the adaptive counts, with
rejection of the chains that place a further photon on an adaptive mode (Sec.~\ref{sec:adaptive});
\item for non-linear sites, where the natural weight fails the lemma, the branch-rejection principle applied to linear
branches, exact with acceptance probability $1/\|c\|_1^2$ (Sec.~\ref{sec:nlbs}).
\end{enumerate}
\noindent Lemma~\ref{lem:chain}, Propositions~\ref{prop:sbs} and~\ref{prop:gap} and the conditioning step of the adaptive
sampler are formalised and machine-checked in Lean~4 (Supplementary Material, Sec.~\ref{S-sec:lean}). All four samplers run in
one GPU code base; they are validated against strong simulation (Sec.~\ref{sec:validation}), and their cost is compared,
measured or modelled, with the single-photon sampler on instances of the same size (Sec.~\ref{sec:table}). What these costs
imply for the size of photonic devices that can be simulated exactly is the subject of a companion study~\cite{Frontier}.

\section{The set-chain lemma}
\label{sec:lemma}

\paragraph{Setting.} An output sample of $k$ detected photons in $m$ modes is drawn as an \emph{ordered} tuple of rows
$z=(z_1,\dots,z_k)\in[m]^k$ from a law $q$, one output mode per photon; forgetting the order gives the occupation pattern
that the detectors record, $s_r=|\{i:z_i=r\}|$ photons in mode $r$. The lemma does not need $q$ to be
exchangeable, $q(z_{\pi(1)},\dots,z_{\pi(k)})=q(z)$ for every permutation $\pi$ (the pattern listed in a uniformly random
order); the laws of this paper are, and Sec.~\ref{sec:adaptive} uses it. There are $N$
\emph{slots} (input columns: photons, sources, emission events), and a non-negative integer \emph{offset} $s_0$: the chain
starts from $s_0$ slots and adds one per placed row, so that after $j$ rows its active set holds $s_j=s_0+j$ slots.

\noindent\emph{Notation:} $[N]=\{1,\dots,N\}$; $A[z,T]$ is the sub-matrix of rows $z$ (with repetition) and columns $T$,
$A[:,t]$ keeps all rows; $z_{<j}=(z_1,\dots,z_{j-1})$; $p!=\prod_lp_l!$ for counts $p$.

\begin{definition}[Chain-consistent set weights]\label{def:weights}
A family $f_j(z;T)\ge0$, defined for $j=0,\dots,k$, $z\in[m]^j$ and $T\subseteq[N]$ with $|T|=s_j$, is \emph{chain-consistent} if
\begin{align}
  & f_0(\emptyset;T)=1 \quad\text{for every } |T|=s_0, \tag{C1}\label{eq:C1}\\
  & \sum_{r\in[m]} f_j(z_1\dots z_{j-1}r;\,T)=\sum_{x\in T} f_{j-1}(z_1\dots z_{j-1};\,T\setminus\{x\})\quad (1\le j\le k). \tag{C2}\label{eq:C2}
\end{align}
Its \emph{target} is $q(z)=\binom{N}{s_k}^{-1}\frac{s_0!}{s_k!}\sum_{|T|=s_k}f_k(z;T)$.
\end{definition}

\begin{lemma}[Set-chain lemma]\label{lem:chain}
Let $f$ be chain-consistent (Definition~\ref{def:weights}), with $s_0+k\le N$. Draw a uniformly random order $\sigma$ of $[N]$; for $j=1,\dots,k$ set $T_j=\{\sigma_1,\dots,\sigma_{s_j}\}$
and draw $z_j$ with probability $f_j(z_1\dots z_{j-1}z_j;T_j)/\sum_r f_j(z_1\dots z_{j-1}r;T_j)$. Then
\begin{enumerate}
\item[(a)] $q$ is a probability law, and its marginal on the first $j$ rows is
$q_j(z_{\le j})=\binom{N}{s_j}^{-1}\frac{s_0!}{s_j!}\sum_{|T|=s_j}f_j(z_{\le j};T)$;
\item[(b)] the algorithm outputs $z$ with law $q$;
\item[(c)] at every step, the law of the active set given the rows drawn so far is
$\Pr[T_j=T\mid z_{\le j}]\propto f_j(z_{\le j};T)$.
\end{enumerate}
\end{lemma}

\begin{proof}
By induction on $j$ with \eqref{eq:C2}, $\sum_{z\in[m]^j}f_j(z;T)=s_j!/s_0!$ for every $T$, so $q$ sums to one. Summing
$q_j$ over the last row and using \eqref{eq:C2}, each $(s_j-1)$-set $T'$ is reached from the $N-s_{j-1}$ sets $T'\cup\{x\}$,
and $\binom{N}{s_j}=\binom{N}{s_{j-1}}\frac{N-s_{j-1}}{s_j}$ gives $\sum_r q_j(z\,r)=q_{j-1}(z)$, which is (a). For (b) and (c)
we show by induction that $\Pr[z_{\le j},T_j=T]=\binom{N}{s_j}^{-1}\frac{s_0!}{s_j!}f_j(z_{\le j};T)$. It holds at $j=0$ by
\eqref{eq:C1}, $T_0$ being a uniform $s_0$-set. The set $T_j$ is $T_{j-1}$ plus a uniform element of its complement, so
\[
\Pr[z_{<j},T_j=T]=\sum_{x\in T}\frac{\Pr[z_{<j},T_{j-1}=T\setminus\{x\}]}{N-s_{j-1}}
=\binom{N}{s_{j-1}}^{-1}\frac{s_0!}{s_{j-1}!}\,\frac{1}{N-s_{j-1}}\sum_{r}f_j(z_{<j}r;T),
\]
the last step by \eqref{eq:C2}. The step probability is therefore needed only where $\sum_rf_j(z_{<j}r;T)>0$: a pair
$(z_{<j},T)$ with a zero denominator is reached with probability zero. Multiplying by the step probability
$f_j(z_{\le j};T)/\sum_rf_j(z_{<j}r;T)$ gives the claim, which yields (c) directly and (b) after summing over $T$ at $j=k$.
\end{proof}

\noindent\emph{Weighted start.} If \eqref{eq:C1} is replaced by $f_0(\emptyset;T)=g(T)\ge0$ and $T_0$ is drawn with
probability $g(T)/W_0$, $W_0=\sum_{|T|=s_0}g(T)$, the remaining slots being added in a uniformly random order, the same
induction gives $\Pr[z_{\le j},T_j=T]=W_0^{-1}\frac{(N-s_j)!}{(N-s_0)!}f_j(z_{\le j};T)$, with unchanged step probabilities;
$g\equiv1$ is the lemma.

\begin{remark}[Scope of the lemma]\label{rem:trivial}
The step weight depends on the \emph{set} $T_j$, not on the order inside it, and the identity \eqref{eq:C2} is used only
after summing over the sets: the pointwise identity $\sum_rf_j(z\,r;T_j)=f_{j-1}(z;T_{j-1})$ along one realised chain is
false already for single photons, so a random order cannot be tested step by step on one chain. Conversely, chain-consistent weights
always exist: $f_j=c_j\,q_j$ with $c_j=s_jc_{j-1}$ satisfies \eqref{eq:C1}--\eqref{eq:C2} for any $q$. The lemma
is therefore a criterion for cheap steps rather than for exact sampling, since every law has a chain rule. A step costs one evaluation of $f_j(z_{<j}r;T_j)$ for all $m$ candidate rows $r$, so the sampler is as fast as the
chain-consistent family it uses. We know no lower bound on the cost of such families, for any of the inputs considered
here.
\end{remark}

\begin{remark}[Fock inputs]\label{rem:fock}
For $n$ single photons in input modes $t$, $A=U[:,t]$, take $N=n$, $s_0=0$ and $f_j(z;T)=|\per A[z,T]|^2$. Condition
\eqref{eq:C2} is the Laplace expansion of the permanent along the last row followed by the orthonormality of the columns of
$A$, and $q(z)=|\per A[z,[n]]|^2/n!$ is the law of the ordered output. This is Clifford and Clifford's
algorithm~B~\cite{CC2018}; the $m$ candidate weights of a step share the $j$ cofactors $\per A[z_{<j},T_j\setminus\{x\}]$,
all obtained from one Glynn sum over $2^{j-1}$ sign vectors~\cite{CC2018}, collapsed to $\prod_\nu(\mu_\nu+1)$ terms when the
placed rows repeat with multiplicities $\mu_\nu$~\cite{CC2020} (Algorithm~\ref{S-alg:fock}, Supplementary Material).
\end{remark}

\begin{figure}[t]\centering
\resizebox{\textwidth}{!}{\begin{tikzpicture}[
    box/.style={rounded corners=3pt, minimum height=1.15cm, align=center, font=\small, inner sep=4pt},
    initbox/.style={box, draw=teal!60!black, fill=teal!8, text width=3.0cm, minimum height=1.5cm},
    chainbox/.style={box, draw=violet!60!black, fill=violet!8, text width=2.9cm},
    stdchain/.style={box, draw=blue!50!black, fill=blue!6, text width=2.9cm},
    title/.style={anchor=south west, inner sep=0pt, font=\small\bfseries},
    panel/.style={draw=black!15, rounded corners=5pt},
    arr/.style={-{Stealth[length=4pt,width=3pt]}, thick, black!60},
    inarr/.style={-{Stealth[length=4pt,width=3pt]}, double, double distance=1.2pt, teal!60!black},
    looparr/.style={-{Stealth[length=4pt,width=3pt]}, red!50!black, thick, rounded corners=4pt},
    looplabel/.style={font=\scriptsize, text=red!50!black, align=center, inner sep=1.5pt, anchor=north},
    chainarr/.style={-{Stealth[length=4pt,width=3pt]}, blue!50!black, thick, rounded corners=3pt},
    legtxt/.style={font=\scriptsize, anchor=west, inner sep=2pt},
  ]

  \def\colsep{8.05}   %
  \def\rowtwo{2.85}\def\rowthree{5.9}   %
  \def\xinit{1.65}    %
  \def\xchain{5.5}    %
  \def\xout{7.6}      %
  \def\xin{0.15}      %
  \def\yrej{-1.22}    %

  \newcommand{\chainloop}[1]{%
    \draw[chainarr] ($(#1.south)+(-0.2,0)$) -- ++(0,-0.35) -- ++(-1.0,0) -- ++(0,0.35);}
  \newcommand{\panelframe}[1][-2.05]{\draw[panel] (-0.15,1.45) rectangle (7.75,#1);}

  \begin{scope}
    \panelframe[-1.2]
    \node[title] at (0,1.0) {Single-photon C\&C\enspace\rtag{\textcolor{black!50}{reference}}};
    \node[stdchain] (C) at (\xchain,0)
      {\textbf{$|\per A[z,T_j]|^2$}\\[1pt]\scriptsize Glynn cofactors};
    \draw[inarr] (\xin,0) -- (C.west);
    \draw[arr] (C.east) -- (\xout,0);
    \chainloop{C}
  \end{scope}

  \begin{scope}[yshift=-\rowtwo cm]
    \panelframe[-1.4]
    \node[title] at (0,1.0) {Superposed sources\enspace\rtag{\textcolor{violet!60!black}{draw $k$ + new weight}}};
    \node[initbox] (I) at (\xinit,0)
      {\textbf{$\cos\alpha|0\rangle{+}\sin\alpha|1\rangle$}\\[1pt]
       \scriptsize draw $k\sim\mathrm{Bin}(n,\sin^2\!\alpha)$,\\ offset $s_0=n-k$};
    \node[chainbox] (C) at (\xchain,0)
      {\textbf{$|\per_\square A[z,T_j]|^2$}\\[1pt]\scriptsize rectangular permanent};
    \draw[inarr] (I) -- (C);
    \draw[arr] (C.east) -- (\xout,0);
    \chainloop{C}
  \end{scope}

  \begin{scope}[xshift=\colsep cm, yshift=-\rowtwo cm]
    \panelframe[-1.4]
    \node[title] at (0,1.0) {Photonic matrix-product state\enspace\rtag{\textcolor{violet!60!black}{new weight}}};
    \node[chainbox] (C) at (\xchain,0)
      {\textbf{$\sum_{e=1}^{R_j}|\langle v_z|\psi_e\rangle|^2$}\\[1pt]
       \scriptsize gapwise Kraus\\ + matrix Glynn};
    \draw[inarr] (\xin,0) -- (C.west);
    \draw[arr] (C.east) -- (\xout,0);
    \chainloop{C}
  \end{scope}

  \begin{scope}[yshift=-\rowthree cm]
    \panelframe
    \node[title] at (0,1.0) {Adaptive optics\enspace\rtag{\textcolor{red!50!black}{prefix init + reject}}};
    \node[initbox, text width=1.45cm] (Ia) at (0.86,0)
      {\textbf{Draw $p$}\\[1pt]\scriptsize PNR counts\\ fix $U^p$};
    \node[initbox, text width=1.45cm] (Ib) at (2.84,0)
      {\textbf{Draw $S$}\\[1pt]\scriptsize prefix set,\\ Lemma\,(c)};
    \node[stdchain] (C) at (\xchain,0)
      {\textbf{$|\per U^p[z,T_j]|^2$}\\[1pt]\scriptsize Fock chain\\ after prefix $a^p$};
    \draw[arr] (Ia) -- (Ib);
    \draw[inarr] (Ib) -- (C);
    \draw[arr] (C.east) -- (\xout,0);
    \chainloop{C}
    \coordinate (R) at ($(C.south)+(1.1,0)$);
    \draw[looparr] (R) -- (R|-0,\yrej) -| (Ib.south);
    \node[looplabel] at ($(Ib.south|-0,\yrej)!0.5!(R|-0,\yrej) + (0,-0.04)$)
      {later photon on an adaptive mode:\\ abort chain, redraw prefix $S$ (same $p$)};
  \end{scope}

  \begin{scope}[xshift=\colsep cm, yshift=-\rowthree cm]
    \panelframe
    \node[title] at (0,1.0) {Non-linear site\enspace\rtag{\textcolor{red!50!black}{propose + reject}}};
    \node[initbox] (I) at (\xinit,0)
      {\textbf{Draw branch $b$}\\[1pt]\scriptsize w.p.\ $|c_b|/\|c\|_1$; linear $U_b$};
    \node[stdchain] (C) at (\xchain,0)
      {\textbf{$|\per U_b[z,T_j]|^2$}\\[1pt]\scriptsize standard C\&C chain};
    \draw[inarr] (I) -- (C);
    \draw[arr] (C.east) -- (\xout,0);
    \chainloop{C}
    \coordinate (A) at ($(C.east)!0.5!(\xout,0)$);
    \fill[red!50!black] (A) circle (1.6pt);
    \draw[looparr] (A) -- (A|-0,\yrej) -| (I.south);
    \node[looplabel] at ($(I.south|-0,\yrej)!0.5!(A|-0,\yrej) + (0,-0.04)$)
      {reject complete sample $s$, redraw $b$:\\ w.p.\ $1-|\textstyle\sum_b c_b a_b(s)|^2/\big(\|c\|_1\sum_b|c_b||a_b(s)|^2\big)$};
  \end{scope}

  \begin{scope}[xshift=\colsep cm]
    \node[title] at (0,1.0) {Legend};
    \node[fill=teal!8, draw=teal!60!black, rounded corners=2pt, minimum width=0.6cm, minimum height=0.36cm]
      (l1) at (0.45, 0.4) {};
    \node[legtxt, right=2pt of l1] {input preparation};
    \node[fill=violet!8, draw=violet!60!black, rounded corners=2pt, minimum width=0.6cm, minimum height=0.36cm]
      (l2) at (0.45, -0.3) {};
    \node[legtxt, right=2pt of l2] {new chain weight};
    \node[fill=blue!6, draw=blue!50!black, rounded corners=2pt, minimum width=0.6cm, minimum height=0.36cm]
      (l3) at (0.45, -1.0) {};
    \node[legtxt, right=2pt of l3] {standard Fock weight};
    \draw[inarr] (4.15, 0.4) -- ++(0.75,0);
    \node[legtxt] at (5.0, 0.4) {random slot order};
    \draw[chainarr] (4.9, -0.17) -- ++(0,-0.26) -- ++(-0.75,0) -- ++(0,0.26);
    \node[legtxt] at (5.0, -0.3) {draw row $r$ (chain step)};
    \draw[looparr, rounded corners=2pt] (4.9, -0.87) -- ++(0,-0.26) -- ++(-0.75,0) -- ++(0,0.26);
    \node[legtxt] at (5.0, -1.0) {rejection};
  \end{scope}

\end{tikzpicture}}
\caption{The samplers of this paper as chains, with the single-photon sampler of Clifford and Clifford as reference. Each
panel shows the input preparation, the weight of the chain (a new weight, or the standard Fock weight), the row-by-row
draw and, for adaptive optics and non-linear sites, the rejection loop: early abort with a new prefix for adaptive
optics, rejection of the complete sample from the branch mixture for a non-linear site. Superposed sources and photonic
MPS construct a weight (Secs.~\ref{sec:sbs}, \ref{sec:ebs}), adaptive optics conditions the Fock chain
(Sec.~\ref{sec:adaptive}), and a non-linear site is sampled by rejection outside it (Sec.~\ref{sec:nlbs}).}\label{fig:samplers}
\end{figure}
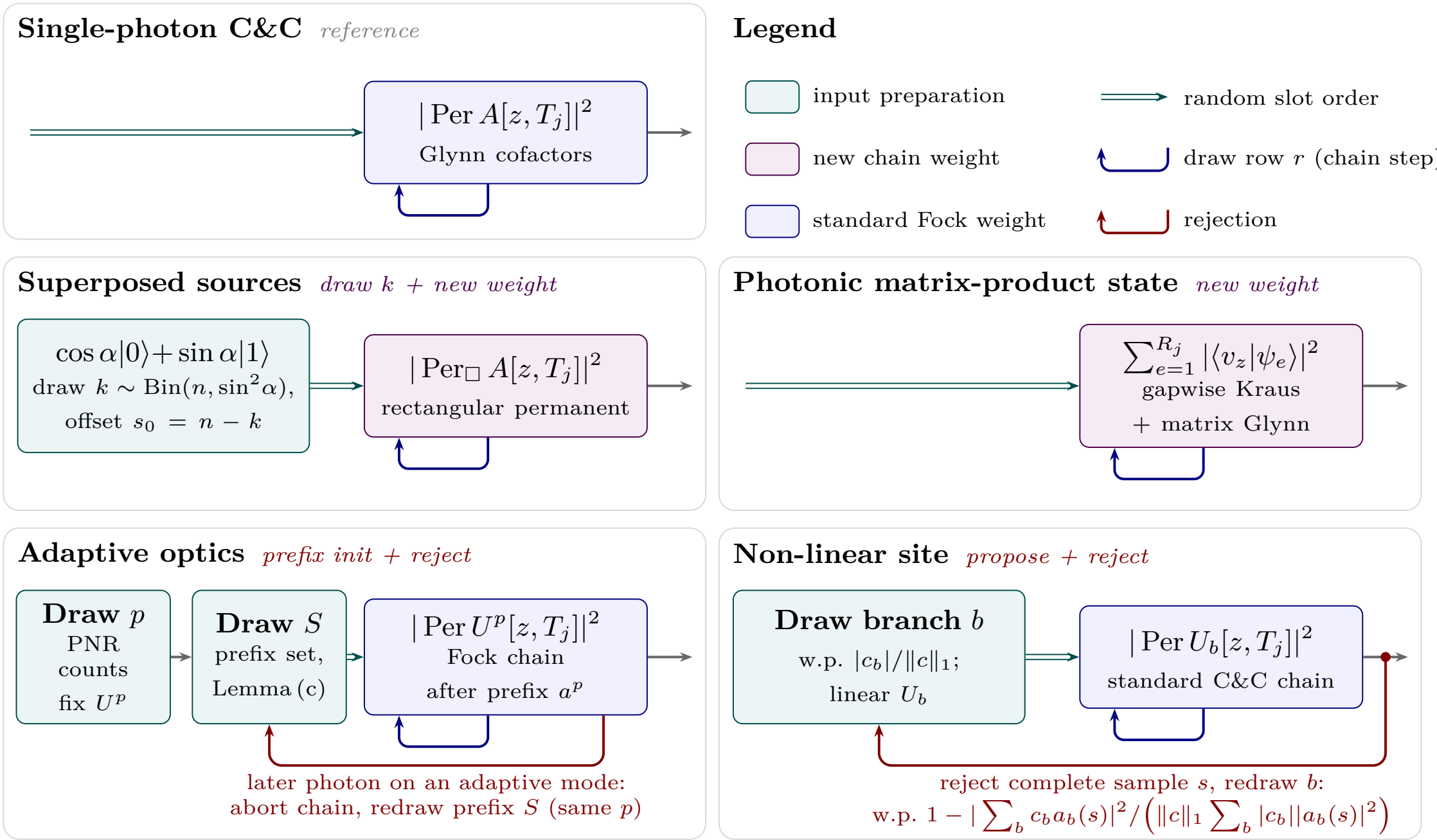

\noindent The lemma is used in three ways in what follows. Sections~\ref{sec:sbs} and~\ref{sec:ebs} \emph{construct a
weight}: one squared amplitude per set for superposed sources, a sum of $R_j$ of them for photonic MPS. Section~\ref{sec:adaptive}
\emph{conditions a chain}: the Fock weight is kept, and part (c) starts the chain from given first rows. Section~\ref{sec:nlbs}
\emph{does not use the lemma}: the natural weight of a non-linear circuit is not chain-consistent, and the sampler is a
rejection from linear branches. Figure~\ref{fig:samplers} shows the four samplers next to the single-photon one.

\paragraph{Loss and partial distinguishability.} Uniform loss, one transmission $\eta$ per photon at the input where it
commutes with the passive interferometer, is a mixture: every sampler first draws which photons survive and then runs on
them, at the cost of the survivors. A lossy superposed source is a mixture of vacuum and a weaker superposed source, and
for a photonic MPS the labels of the lost emissions are drawn first, which keeps the state of the kept emissions pure.
Partial distinguishability is treated in the mixture ($x$-)model, each photon or source independently in a common internal
mode with probability $x$ and otherwise in an orthogonal mode of its own. For product inputs (single photons, superposed
sources) a photon in its own mode is routed independently. For the other three samplers it is not independent of the
rest and enters a joint draw: for a photonic MPS its output row is drawn with the lost labels and its slot becomes a
fixed operator of the train; in an adaptive circuit it is drawn level by level inside the feed-forward, whose counts it
changes; at a non-linear site it shares the branch index, and its amplitude enters the coherent branch sum. Details, and
the cases not covered (a general Gram matrix of internal states, loss between two non-linear sites, inefficient adaptive
detectors), are in the Supplementary Material, Sec.~\ref{S-sec:imperfections}.

\section{Superposed sources}
\label{sec:sbs}

\paragraph{Model.} Each of $n$ sources emits the same state $\cos\alpha\ket0+\sin\alpha\ket1$, source $i$ into input mode
$t_i$ (unequal sources at the end of this section); the photons pass through an $m$-mode interferometer $U$ and are
detected with photon-number resolution~\cite{Renema2020SBS}. A passive interferometer conserves photon
number, so the detected count is exactly $k\sim\mathrm{Bin}(n,\sin^2\alpha)$ whatever $U$, and given $k$ the input is the
Dicke state $\binom nk^{-1/2}\sum_{|S|=k}\prod_{i\in S}a^\dagger_{t_i}\ket0$. With $A=U[:,t]$ ($m\times n$, orthonormal
columns), the amplitude of an ordered $k$-row output $z$ is the \emph{rectangular permanent}
\[
  \perrect A[z,[n]] = \sum_{f:[k]\hookrightarrow[n]}\ \prod_{t=1}^k A_{z_t f(t)} = \sum_{|S|=k}\per A[z,S],
\]
and $q(z)=\frac{(n-k)!}{n!}\,|\perrect A[z,[n]]|^2$. Here, as throughout, $q$ is a law on ordered tuples, collisions
included: an occupation pattern $s$ with $|s|=k$ has probability $\frac{k!}{\prod_is_i!}\,q(z)$ for any ordering $z$ of $s$. At $\alpha=\pi/4$ the mean count is $n/2$ and the binomial upper tail
carries the cost. The incoherent mixture of the same sources --- each source firing with probability $\sin^2\alpha$,
which source fired being recorded --- has the same photon-number law; it is single photons under input loss
$\eta=\sin^2\alpha$.

\paragraph{The offset-start order.} A random order over the \emph{sources}, not the photons, gives chain-consistent weights
once the active set starts with $n-k$ sources instead of none: after the $k$ steps it then holds all $n$, and the weight
of a set is the rectangular permanent of the rows placed so far.
\begin{proposition}\label{prop:sbs}
For superposed sources conditioned on $k$ photons, $N=n$, $s_0=n-k$ and
$f_j(z;T)=|\perrect A[z,T]|^2$ ($j$ rows, $n-k+j$ columns) are chain-consistent weights with target $q$.
\end{proposition}
\begin{proof}
\eqref{eq:C1}: the empty rectangular permanent is $1$. \eqref{eq:C2}: expand $|\perrect A[z_{<j}r,T]|^2$ over the column
$x$ used by the last row in each of the two injections; summing over $r$, $\sum_rA_{rx}\bar A_{rx'}=\delta_{xx'}$ removes
the cross terms and leaves $\sum_{x\in T}|\perrect A[z_{<j},T\setminus\{x\}]|^2$. Target: at $j=k$ the only set is $T=[n]$,
and $\binom{n}{n}^{-1}\frac{(n-k)!}{n!}f_k(z;[n])=q(z)$.
\end{proof}

\noindent At $k=n$ the offset vanishes and the chain is Clifford and Clifford's. At $k<n$ the $n-k$ sources that supply no
photon are never identified. Part (a) of the lemma recovers the true marginal
$q_j(z)\propto\sum_{|R|=k-j}|\perrect A[z,[n]\setminus R]|^2$, the residual-set form that a direct sampler would have to
enumerate.

\paragraph{The step.} With $j-1$ rows placed, the $m$ candidate weights share the cofactors
$P_x=\perrect A[z_{<j},T_j\setminus\{x\}]$, $x\in T_j$, and $w(r)=|\sum_{x\in T_j}A_{rx}P_x|^2$. Glynn's identity~\cite{Glynn2010} with one
sign per placed row gives every cofactor at once,
\[
  P_x = 2^{-(j-1)}\sum_{\delta\in\{\pm1\}^{j-1}}\Big(\prod_t\delta_t\Big)\;e_{j-1}\big(a(\delta)\setminus a_x(\delta)\big),\qquad
  a_i(\delta)=\sum_t\delta_tA_{z_ti},
\]
$e_{j-1}$ the elementary symmetric polynomial of degree $j-1$ of the $|T_j|-1$ values other than $a_x$. Since $e_{j-1}$ is
homogeneous, the global flip $\delta\to-\delta$ leaves the summand invariant and $2^{j-2}$ signs suffice. The leave-one-out
value $e_{j-1}(a\setminus a_x)$ is the coefficient of $\lambda^{j-1}$ in $\prod_{i\ne x}(1+\lambda a_i)$ and, since the
offset is constant along the chain, $|T_j|-j=s_0$, also the coefficient of $\mu^{s_0}$ in $\prod_{i\ne x}(\mu+a_i)$. Prefix
and suffix products of the linear factors, truncated at degree $d=\min(j-1,s_0)$ in whichever variable is shorter, give every
leave-one-out value without division in $O(|T_j|(d+1))$; a cheaper division, stable in one direction, is described in
Sec.~\ref{S-sec:division}. At $s_0=0$ this is the all-but-one product of the single-photon step, so at $k=n$
the chain is Clifford and Clifford's step for step, payload included. The collapse of repeated rows~\cite{CC2020} applies
unchanged to the sign sum: placed rows repeated with multiplicities $\mu_\nu$ contribute $\prod_\nu(\mu_\nu+1)$ terms instead
of $2^{j-1}$, walked in the same reflected Gray order as for single photons. The cost of a step is
$O(\prod_\nu(\mu_\nu+1)\,|T_j|(d+1)+m|T_j|)$ and of a sample at most
\[
  C(k)=O\big(2^k n^2+mnk\big),\qquad \E_k[C]\le O\big(\E[2^k]\,n^2+mn^2\big)=O\big((1+\sin^2\alpha)^n n^2+mn^2\big),
\]
the exponent of single photons under loss $\eta=\sin^2\alpha$; the collapse lowers the term count as it does for single photons~\cite{CC2020}.
Loss and partial distinguishability are drawn before the chain (Sec.~\ref{sec:lemma}). Algorithm~\ref{S-alg:sbs} (Supplementary
Material); validation in Sec.~\ref{sec:validation}.

\paragraph{Unequal sources.} Proposition~\ref{prop:sbs} extends to sources $a_i\ket0+b_i\ket1$ with different amplitudes,
$a_i\ne0$. With $u_i=b_i/a_i$ and the scaled columns $\tilde A=U[:,t]\,\mathrm{diag}(u)$, the output amplitude given $k$ photons is
$\perrect\tilde A[z,[n]]$ up to a constant, and the weights $f_j(z;T)=g(T)\,|\perrect\tilde A[z,T]|^2$, $g(T)=\prod_{i\in T}|u_i|^{-2}$,
satisfy \eqref{eq:C2}: expanding the last row now leaves $\sum_{x\in T}|u_x|^2|\perrect\tilde A[z_{<j},T\setminus\{x\}]|^2$,
and $g(T)|u_x|^2=g(T\setminus\{x\})$. With the weighted start of Sec.~\ref{sec:lemma}, $T_0$ drawn among the $(n-k)$-sets
with probability $\propto\prod_{i\in T_0}|u_i|^{-2}$ (a polynomial draw by elementary symmetric polynomials), the chain
samples the output law conditioned on $k$, whose photon number is the Poisson-binomial law of the $|b_i|^2$. The factor $g$
cancels in every step, so a step is that of equal sources on the columns of $\tilde A$: dim sources are more likely to start
in the unused set. A source with $b_i=0$ never fires and is removed; one with $a_i=0$ always fires and never enters $T_0$.
An exact enumeration of this chain over initial sets, orders and output rows reproduces the output law of five random
instances with unequal sources, $n\le5$, to $2\times10^{-16}$; our GPU implementation covers equal sources.

\section{Photonic matrix-product states}
\label{sec:ebs}

\paragraph{Model.} A quantum emitter with an internal memory of dimension $\chi$ (a spin, $\chi=2$; for instance a
quantum-dot spin driven to emit linear graph states~\cite{Huet2025}) emits $n$ photons one after the other;
emission $s$ puts one photon into one of $L$ input modes $t_{s,l}$ with operator $M^{(s)}_l$ ($\chi\times\chi$), and the
photonic state is the matrix-product state~\cite{Schon2005}
\[
  \ket\psi=\sum_{x\in[L]^n}\bra{\mathrm{out}}M^{(n)}_{x_n}\cdots M^{(1)}_{x_1}\ket{\mathrm{in}}\,\ket{x},\qquad
  \ket x=\prod_s a^\dagger_{t_{s,x_s}}\ket0,
\]
with $\ket{\mathrm{in}}$ and $\bra{\mathrm{out}}$ the initial and final emitter states and all $nL$ input modes distinct
(dual-rail time bins for $L=2$); a traced final emitter state is a first draw. The
caterpillar (spin-photon cluster) states~\cite{Lindner2009} are $M^{(s)}_l=\ket l\!\bra l G_s$ with spin gates $G_s$.
The model contains single-photon boson sampling ($\chi=1$, $L=1$, $M^{(s)}_1=1$), so exact sampling from it is at least as
hard, in the worst case, as from Fock inputs~\cite{AA2013}; what is known for a fixed train is summarised in the
Supplementary Material, Sec.~\ref{S-sec:hardness}.

\noindent We assume $\braket{\psi|\psi}=1$: the emissions are isometric, $\sum_lM^{(s)\dagger}_lM^{(s)}_l=1$, and a traced
final emitter state is drawn first, with probability $p_{\rm out}$; the displayed amplitudes are then divided by $\sqrt{p_{\rm out}}$,
$\psi(x)=\bra{\mathrm{out}}W_x\ket{\mathrm{in}}/\sqrt{p_{\rm out}}$ with $W_x=M^{(n)}_{x_n}\cdots M^{(1)}_{x_1}$, or equivalently $\bra{\mathrm{out}}$ is rescaled.

\paragraph{Weights.} The slots are the emissions, $N=n$, $s_0=0$. For a set $T$ of emissions let
$\rho_T=\Tr_{T^c}\ket\psi\!\bra\psi$, a state of $|T|$ photons, and for $z\in[m]^{|T|}$ let $\bra{v_z}$ be the covector
\[
  \braket{v_z|\phi}=\sum_{x_T}\per U[z,t_{T,x_T}]\,\phi(x_T),
\]
the amplitude of the ordered output $z$ for a $|T|$-photon state $\phi$ on the modes $t_T$; the components of the ket
$\ket{v_z}$ are the conjugates $\overline{\per U[z,t_{T,x_T}]}$. Then $f_j(z;T)=\bra{v_z}\rho_T\ket{v_z}$
is chain-consistent: \eqref{eq:C1} is the trace; for \eqref{eq:C2}, expand both permanents of $f_j(z_{<j}r;T)$ along the last row,
$\per U[z_{<j}r,t_{T,x_T}]=\sum_{c\in T}U_{r,t_{c,x_c}}\per U[z_{<j},t_{T\setminus c,x_{T\setminus c}}]$; the sum over $r$ gives
$\sum_rU_{r,t_{c,x_c}}\bar U_{r,t_{c',y_{c'}}}=\delta_{cc'}\delta_{x_cy_c}$ because the $nL$ input modes are distinct, and
$\sum_{x_c}\rho_T(x_T,y_T)|_{y_c=x_c}=\rho_{T\setminus c}$ leaves $\sum_{c\in T}f_{j-1}(z_{<j};T\setminus c)$. The target is the
output law of $\psi$. The order must be uniformly random: with a fixed emission order the weights are not the marginals of
one law (TVD 0.22 against a floor of 0.05 on a three-photon linear cluster).

\paragraph{Gapwise Kraus contraction.} $\rho_T$ is mixed because the emissions outside $T$ are traced, and for a random
order $T$ is an arbitrary subset of the train: its complement splits into disjoint maximal runs of traced emissions, the
\emph{gaps}. A contraction of $\rho_T$ with the transfer map $X\mapsto\sum_lM_lXM_l^\dagger$ handles a gap in time linear in
its length --- it is how a matrix-product state is sampled site by site with a chain rule over its own
modes~\cite[\S6.2.4]{GPRS2019} --- but it yields a density matrix, not the sum of squared amplitudes that our chain over
output rows and Glynn's sum need; a
pure-state unravelling of the traced labels would enumerate the $L^{|g|}$ label strings of every gap $g$. The following
proposition compresses each gap's transfer map into at most $\chi^2$ Kraus operators, whatever its length, and so keeps a
sum of squared amplitudes with few terms.

\begin{proposition}[Gapwise Kraus contraction]\label{prop:gap}
Let $T=\{c_1<\dots<c_j\}$, and for a gap $g$ (a maximal run of emissions not in $T$) let $A^{(g)}_{x_g}$ be the ordered
product of its emission operators and $E_g(X)=\sum_{x_g}A^{(g)}_{x_g}XA^{(g)\dagger}_{x_g}$ its map on $\chi\times\chi$
matrices. Choose any Kraus decomposition $E_g(X)=\sum_iQ_{g,i}XQ_{g,i}^\dagger$ of every inner gap (between $c_1$ and $c_j$),
any vectors with $\sum_xA_x\ket{\mathrm{in}}\!\bra{\mathrm{in}}A_x^\dagger=\sum_i\ket{\xi_i}\!\bra{\xi_i}$ for the prefix gap (before $c_1$) and any
covectors with $\sum_xA_x^\dagger\ket{\mathrm{out}}\!\bra{\mathrm{out}}A_x=\sum_i\ket{\phi_i}\!\bra{\phi_i}$ for the suffix gap (after $c_j$). For a
choice $e$ of one index per gap let
\[
  \psi_e(x_T)=\bra{\phi_{i_{\rm suf}}}M^{(c_j)}_{x_{c_j}}Q_{g_{j-1},i_{j-1}}\cdots Q_{g_1,i_1}M^{(c_1)}_{x_{c_1}}\ket{\xi_{i_{\rm pre}}}.
\]
(an absent gap contributes the identity, or $\ket{\mathrm{in}}$, $\bra{\mathrm{out}}$ at the ends). Then $\rho_T=\sum_e\ket{\psi_e}\!\bra{\psi_e}$,
and every $\psi_e$ is a matrix-product state of bond dimension $\chi$ on the emissions of $T$. Inner gaps have Kraus rank at most $\chi^2$ and end gaps at most $\chi$, so the number of terms is
$R_j\le\min\big(L^{n-j},\chi^{2g_{\rm in}+g_{\rm end}}\big)$, with $g_{\rm in}$ inner and $g_{\rm end}\le2$ end gaps.
\end{proposition}
\begin{proof}
$\rho_T(x_T,y_T)=\sum_{x_{T^c}}\psi(x_Tx_{T^c})\overline{\psi(y_Tx_{T^c})}$. Read $\psi(x)\overline{\psi(y)}$ as
$\bra{\mathrm{out}}W_x\ket{\mathrm{in}}\!\bra{\mathrm{in}}W_y^\dagger\ket{\mathrm{out}}$ with $W_x=M^{(n)}_{x_n}\cdots M^{(1)}_{x_1}$, and contract from the right:
starting from $X=\ket{\mathrm{in}}\!\bra{\mathrm{in}}$, a kept emission $c$ maps $X\mapsto M^{(c)}_{x_c}XM^{(c)\dagger}_{y_c}$, and the labels of
a gap occur only in that gap's factors, once on each side, so summing them maps $X\mapsto E_g(X)$. Each $E_g$ may be replaced
by any of its Kraus decompositions, the prefix gap by the vectors $\ket{\xi_i}$ and the suffix gap by the covectors
$\bra{\phi_i}$; expanding the products of sums gives $\sum_e\psi_e(x_T)\overline{\psi_e(y_T)}$. The Choi matrix of $E_g$ is
$\chi^2\times\chi^2$, which bounds an inner gap's Kraus rank by $\chi^2$; the end gaps produce positive $\chi\times\chi$
matrices, rank at most $\chi$. The bound $L^{n-j}$ is the explicit sum over the traced labels.
\end{proof}

\noindent We take the minimal decompositions, by an SVD of the stacked $A^{(g)}_{x_g}$ reduced emission by emission, so the
cost of a gap is linear in its length. For instance, with $n=3$ and $T=\{1,3\}$, the middle emission is a gap of length one,
$E(X)=\sum_lM^{(2)}_lXM^{(2)\dagger}_l$, and $\psi_e(x_1,x_3)=\bra{\mathrm{out}}M^{(3)}_{x_3}Q_eM^{(1)}_{x_1}\ket{\mathrm{in}}$ with
$e\le\min(L,\chi^2)$ terms.

\paragraph{Matrix-valued Glynn sums.} So $f_j=\sum_e|\braket{v_z|\psi_e}|^2$, a sum of $R_j$ squared amplitudes, and each
amplitude factorises through the matrix product inside Glynn's sum. With column sums
$s_{c,l}(\delta)=\sum_i\delta_iU_{z_i,t_{c,l}}$ and $G_c(\delta)=\sum_ls_{c,l}(\delta)\tilde M^{(c)}_l$, $\tilde M^{(c)}_l$ the
kept emission's operator with the Kraus operator of the gap before it absorbed,
\[
  \braket{v_z|\psi_e}=2^{-(j-1)}\sum_{\delta\in\{+1\}\times\{\pm1\}^{j-1}}\Big(\prod_i\delta_i\Big)\,\bra{\mathrm{out}_e}G_{c_j}(\delta)\cdots G_{c_1}(\delta)\ket{\mathrm{in}_e},
\]
because Glynn's formula $\per B=2^{-(j-1)}\sum_{\delta_1=1}(\prod_i\delta_i)\prod_c(\sum_i\delta_iB_{ic})$, applied to
$B_{ic}=U_{z_i,t_{c,x_c}}$, is multilinear in the columns and the sum over $x_T$ passes inside the product. For the Laplace
step the last row is left out: the candidate amplitude is
$\sum_{c\in T}\sum_lU_{r,t_{c,l}}\Tr\big(\tilde M^{(c)}_lK_c\big)$ with the \emph{matrix} cofactors
\begin{gather*}
  K_c=2^{-(j-2)}\sum_{\delta\in\{+1\}\times\{\pm1\}^{j-2}}\Big(\prod_i\delta_i\Big)\,p_c(\delta)\,q_c(\delta)^{T},\\
  p_c=G_{c-1}\cdots G_{c_1}\ket{\mathrm{in}_e},\qquad q_c^T=\bra{\mathrm{out}_e}G_{c_j}\cdots G_{c+1},
\end{gather*}
where $c-1$ and $c+1$ denote the kept emissions before and after $c$ in $T$ (not the emissions $c\mp1$), and the Glynn sum runs over the $j-1$ placed rows (the global flip leaves the summand of a $(j-1)\times(j-1)$ permanent invariant) ---
the scalar prefix and suffix products of the Fock kernel replaced by $\chi\times\chi$ ones. For $j\ge2$; at $j=1$ no row is
placed and $K_{c_1}=\ket{\mathrm{in}_e}\bra{\mathrm{out}_e}$. The 2020 collapse acts on $\delta$
only and carries over unchanged. Algorithm~\ref{S-alg:ebs} (Supplementary Material).

\paragraph{The term count.} With $C_j$ Glynn terms ($2^{j-2}$, fewer after the collapse), a step costs
$O\big(R_j\,(C_j\,jL\chi^2+mjL)\big)$: per term and sign vector, $j$ matrices $G_c$ and the prefix and suffix sweeps, then
the $m$ candidates. Measured means of $R_j$ are far below the bound of Proposition~\ref{prop:gap}, peaking at $j\approx n/2.5$ (7 at $n=8$, 30 at $n=12$, 96
at $n=16$ on a linear cluster) and falling to 1--2 on the last steps, which carry the $2^j$ weight. The per-gap
decomposition is minimal gap by gap but not globally; comparing $R_j$ with the rank of $\rho_T$ over all steps, the surplus
is up to $16\times$ on single early steps but, weighted by step cost, $1.02$--$1.11\times$ for $n=8$--$20$. Since the rank is
the smallest number of terms any decomposition of $\rho_T$ can have, a better decomposition would save at most about
$10\%$ of the run time. %

\section{Adaptive linear optics}
\label{sec:adaptive}

\paragraph{Model.} $n$ single photons enter $U_0$; $K$ output modes $a_1..a_K$ are read with photon-number resolution in
$\Lambda$ levels, and after each level a classical function of the counts $p_{\le\lambda}$ read so far chooses the unitary
$V_\lambda$ applied to the unmeasured modes~\cite{CMS2021}. Each adaptive outcome $p$ has its cumulative unitary $U^p$ and
$P(p,s)=|\per U^p[(p,s),t]|^2/(p!s!)$, with $(p,s)$ the output occupations read as a tuple of rows and $p!$, $s!$ the
products of the factorials of the counts. Oh and Lim~\cite{OhLim2025adaptive} show that mean values of a restricted class of
observables are estimable in polynomial time for $K=O(1)$ and BQP-complete for $K=\mathrm{poly}(n)$, and sample exactly and efficiently when the input and the final
measurement are Gaussian and $K=O(\log m)$; for Fock input with photon-number-resolved final detection they give marginals
and mean values, not a sampler.

\paragraph{Stage 1: the adaptive outcome.} $P(p_{\le\lambda})$ depends on $U^{p_{<\lambda}}$ only, since later $V$ act on
modes that are traced in it; the outcomes form a chain over levels whose conditionals are $K_\lambda$-mode marginals of a
non-adaptive sampler. We compute them exactly with the generating-function algorithm of Ivanov and Gurvits~\cite{IvanovGurvits2020}, as used by
Oh and Lim~\cite{OhLim2025adaptive}: the generating
function of the counts on a set $\mathcal A$ of adaptive modes is $\per(1+V^\dagger(Z-1)V)$, $V=U[\mathcal A,t]$,
$Z=\mathrm{diag}(\zeta_a)_{a\in\mathcal A}$ formal variables, whose coefficients are the binomial moments $M[p]$,
accumulated photon by photon on a table graded by total degree and inverted. This fits for $K\le4$ at $n=38$. Beyond, the
counts of level $\lambda$ can be read from a stage-2 chain on $U^{p_{<\lambda}}$ conditioned on the earlier counts
$p_{<\lambda}$ and accepted with probability $P(p_{<\lambda})/M[p_{<\lambda}]$; the later unitaries act on unmeasured modes,
so the draw is exact. We did not implement this case.

\paragraph{Stage 2: conditioning by the lemma.} The final counts given $p$ are the Fock output of $U^p$ conditioned on ``the
adaptive modes show exactly $p$''. By exchangeability of the ordered law this is the conditional given ``the first $r=|p|$
rows are $a^p$ (the adaptive modes with multiplicity) and no later row is adaptive''. Part (c) of the lemma turns the first
half into a draw: the active set after $r$ rows has law $\propto|\per U^p[a^p,S]|^2$. Because $a^p$ has only $K$ distinct
rows, a permanent with repeated rows is a coefficient of a product of linear forms,
$\per U^p[a^p,S]=p!\,[x^p]\prod_{j\in S}(B_j\cdot x)$ with $B_j=U^p[a,t_j]\in\C^K$ and $p!=\prod_lp_l!$, and deciding the
columns one by one from suffix tables of $\prod_{i\ge j}(1+(B_i\cdot x)(\bar B_i\cdot w))$ draws $S$ from this posterior
with tables of $\prod_l(p_l+1)^2$ entries, polynomial for fixed $K$ (proof in the Supplementary Material, Sec.~\ref{S-sec:columns}). The
chain then continues with ordinary Clifford--Clifford steps.

\paragraph{The remaining constraint.} The second half --- no later photon on an adaptive mode --- is \emph{not} a
chain-consistent weight: restricting the rows to the non-adaptive modes makes the columns of the interferometer
sub-orthonormal, and the Gram matrix $1-V^\dagger V$ then couples different hidden sets, so \eqref{eq:C2} fails. It is
enforced by rejection with early abort: a chain is
discarded at the first photon that lands on an adaptive mode. A chain is accepted with probability $P(p)/M[p]$, $M$ the binomial moments: by exchangeability, $P(p)$ and $M[p]$ are the
probabilities that the first $r$ rows are $a^p$ with and without the constraint, each multiplied by the same factor
$n!/((n-r)!\,p!)$.

\paragraph{Cost.} A branch $p$ starts $M_{U^p}[p]/P(p)$ chains, $M_{U^p}$ the binomial moments of its own interferometer,
so the mean number of chains started per sample is $\sum_{p:\,P(p)>0}M_{U^p}[p]$, over the outcomes that occur. For a fixed
interferometer, $U^p=U$, and when every outcome has positive probability, the generating function of the binomial moments
turns this into $\E[2^r]=\per(1+V^\dagger V)$, $r=|p|$ the number of photons the adaptive modes receive; with feed-forward
the moments of different outcomes come from different interferometers and no such closed form holds. Most rejected
chains stop early: given the forced prefix, the remaining $n-r$ rows are exchangeable, so a chain with $h\ge1$ further photons on the adaptive modes places them at a uniformly random $h$-subset of
its free steps and stops at the first, $J$, at step $(n-r+1)/(h+1)$ on average; with a step cost growing like $\varrho^j$, it
is charged $(\varrho^{r+J}-\varrho^r)/(\varrho^n-1)$ of a full chain. A cost model built on the first-level interferometer, at $m=2n$
and $n=38$, gives $\E[2^r]=2.0$, $3.8$, $6.8$ chains for $K=1,2,3$ and a time per sample $X=1.03$, $1.07$, $1.10$ times the
plain sampler's; with a Haar policy we measure $2.1$ and $4.1$ chains per sample on average for $K=1,2$ at
$n=30$--$38$ ($1.8$--$2.3$ and $3.5$--$4.7$ per point, $256$ samples each). %
Algorithms~\ref{S-alg:adaptive} and~\ref{S-alg:prefix} (Supplementary Material).

\section{Non-linear sites}
\label{sec:nlbs}

\paragraph{Model.} A single-mode phase $e^{-if(\hat n_x)}$ acts on mode $x$ between Haar interferometers $U_0$ and $U_1$; the
response is a Kerr phase, $f(j)=\varphi j^2$~\cite{Spagnolo2021}, or any response diagonal in the photon number. On $n$ photons $\hat n_x$ has
eigenvalues $0..n$, so the response is exactly a finite sum of phase shifters,
$e^{-if(\hat n_x)}=\sum_\ell c_\ell e^{i\theta_\ell\hat n_x}$, and the circuit is a coherent sum of $B$ linear interferometers
$U_b=U_1D_x(\theta_b)U_0$ with weights $c_b$. The decomposition is not unique; we use the shortest periodic one: the
response $e^{-if(j)}$ on $j=0..n$ is a linear phase $e^{iaj}$ times a function of smallest period $q\le n$, expanded on its
$q$-point discrete Fourier transform, $\theta_\ell=a+2\pi\ell/q$, so that $B\le q$ and $\|c\|_2=1$.
$B=1$ at $\varphi=0$ and $\varphi=\pi$, $2$ at $\pi/2$, $4$ at $\pi/4$, and
$n$ for a response without short period, such as $\varphi=1$; several sites multiply the branches.

\paragraph{Failure of (C2).} For a linear circuit the traced photons contribute $\per G[T^c,S^c]=\delta_{ST}$ with
$G=U^\dagger U=1$ (Remark~\ref{rem:fock}), so the marginal is diagonal in the hidden set. Here two branches meet through
$G_{bb'}=U_{b'}^\dagger U_b=1+(e^{i(\theta_b-\theta_{b'})}-1)\,ww^\dagger$, $w=U_0^\dagger e_x$, whose rank-one term couples
different hidden sets. The natural set weight $f_j(z;T)=|\sum_bc_b\per U_b[z,T]|^2$ therefore violates \eqref{eq:C2}: the
random-order chain built on it misses the law by a TVD of $0.115$ at $n=3$, $m=4$, Kerr $0.9$, where the same enumeration
reproduces a linear circuit to $10^{-15}$. The marginal, the cost of the true-marginal chain and the purifications we tried
are in the Supplementary Material, Sec.~\ref{S-sec:nlbs-fail}.

\paragraph{Rejection from the branch mixture.} Draw a branch $b$ with probability $|c_b|/\|c\|_1$, an ordinary Fock sample
$s$ of $U_b$ (the Fock kernels unchanged), and accept with probability
$|\sum_bc_ba_b(s)|^2/(\|c\|_1\sum_b|c_b||a_b(s)|^2)$, $a_b(s)=\per U_b[s,t]/\sqrt{s!}$ the amplitude of the occupation
$s$. By Cauchy--Schwarz, $|\sum_bc_ba_b|^2\le\|c\|_1\sum_b|c_b||a_b|^2$, so this is a probability. The proposal draws $s$ with
probability $\sum_b(|c_b|/\|c\|_1)|a_b(s)|^2$, so $s$ is proposed and accepted with probability
$|\sum_bc_ba_b(s)|^2/\|c\|_1^2$: proportional to the target, and summing to $1/\|c\|_1^2$ because
$\sum_s|\sum_bc_ba_b(s)|^2=\|\sum_bc_bU_b\ket{t}\|^2=1$, the circuit being unitary. The sampler is exact, with acceptance
probability exactly $1/\|c\|_1^2$ (measured 0.43--0.59 against 0.50 at $\|c\|_1^2=2$; 0.054--0.077 against 0.0625 at $B=16$). An accepted
sample costs $(1+\kappa B)\|c\|_1^2$ Fock samples, $\kappa=0.55\pm0.08$ the measured cost of one branch permanent. For a
fixed number of sites and a response of bounded period ($B=O(1)$) this is a constant factor and the exponential rate in $n$
is that of single photons; for a generic response $B=O(n)$ and the factor is polynomial. Algorithm~\ref{S-alg:nlbs}
(Supplementary Material). This is the $|c_b|$-weighted specialisation of the rejection principle of Dias and
K\"onig~\cite{DiasKoenig2025}, applied to the branch decomposition with weights $|c_b|$ instead of $|c_b|^2$: by Cauchy--Schwarz $\|c\|_1^2\le B\|c\|_2^2$, and for a given decomposition the weights $|c_b|$ minimise the constant (at $\varphi=\pi/2$ the
two coincide, $\|c\|_1^2=B=2$; at $\varphi=1$, $n=40$, $33.2$ against $40$). The constant $\|c\|_1^2$ depends on the
decomposition: another set of phase shifters, not restricted to the discrete Fourier frequencies, can lower it, possibly
with more branches; all the numbers here use the shortest periodic one. Spagnolo et al.'s exact simulation
replaces the site by a linear-optical gadget with ancillary photons and postselection~\cite{Spagnolo2021}; the branch form
needs no ancilla, and its acceptance probability is known in closed form.

\paragraph{Weak non-linearities and several sites.} For a fixed $n$ the rejection factor $\|c\|_1^2$ tends to one as
$\varphi\to0$, but it is governed by $\varphi n^2$ rather than $\varphi$. A Kerr phase $\varphi=0.01$ already costs $\|c\|_1^2=8.6$ at
$n=40$, while $\varphi=\pi/2$ costs $2$. Over $L$ sites at $\varphi=\pi/2$ the rejection factor is $2^L$, and each proposal
evaluates $2^L$ branch amplitudes, about $4^L$ in all, so the exponent changes once the number of sites grows with $n$. A
chain-consistent family for $\sum_bc_bU_b$ with polynomially many terms per set would avoid the rejection and make a weak
non-linearity cheap; we know none, and no lower bound that would exclude one.

\section{Comparison of the samplers}
\label{sec:compare}

Table~\ref{tab:criterion} compares the samplers and evaluates their step cost and overhead over single photons.

\begin{table}[h]\centering\footnotesize
\setlength{\tabcolsep}{4pt}\begin{tabular}{L{1.9cm}L{1.5cm}L{0.7cm}L{2.7cm}L{0.8cm}L{2.2cm}L{2.75cm}}\toprule
input / circuit & slots & $s_0$ & weight $f_j(z;T)$ & terms & step cost & overhead over single photons\\\midrule
single photons & photons & 0 & $|\per A[z,T]|^2$ & 1 & $2^{j}j+mj$ & ---\\
superposed sources & sources & $n-k$ & $|\perrect A[z,T]|^2$ & 1 & $2^{j}|T|(d+1)+m|T|$ & payload factor $|T|(d+1)/j$, $d=\min(j-1,s_0)$\\
photonic MPS & emissions & 0 & $\bra{v_z}\rho_T\ket{v_z}$ & $R_j$ & $R_j(2^{j}jL\chi^2+mjL)$ & term count $R_j$\\
adaptive optics & photons & 0 & $|\per U^p[z,T]|^2$, started by (c) & 1 & $2^{j}j+mj$ & rejection of chains that put a further photon on an adaptive mode: $\sum_pM_{U^p}[p]$ chains\\
non-linear site & photons & --- & none known & --- & $2^{j}j+mj$ per branch sample & rejection, $(1+\kappa B)\|c\|_1^2$\\\midrule
any input, generic chain~\cite{Renema2021marginals} & photons & 0 & the true marginal $c_jq_j$ & --- & (input pairs) $\times$ subsets & none of its own; no speed-up\\
any superposition~\cite{DiasKoenig2025} & --- & --- & none (rejection) & --- & one component sample and the target amplitude & mixture of $N_{\rm c}$ components: $N_{\rm c}\|\beta\|_2^2$\\\bottomrule
\end{tabular}
\caption{How the four methods use --- or depart from --- the set chain; step costs up to constants, $j$ placed photons, $m$
candidate rows, before the collision collapse of Ref.~\cite{CC2020}. Loss is drawn before the chain in every row;
distinguishability as in Sec.~\ref{sec:lemma}.}\label{tab:criterion}
\end{table}

\section{Numerical evaluation}
\label{sec:numerics}

\subsection{Validation}
\label{sec:validation}

Every sampler is checked on small instances, within reach of strong simulation, through the same call and GPU backend as
the cost measurements of Sec.~\ref{sec:table}. The statistic is the total variation distance (TVD) between the empirical
distribution of $S=4\cdot10^4$ samples and the oracle's. Its reference is the TVD of an exact draw of the same size: we
draw $40$ multinomial samples of size $S$ from the oracle's distribution and take the mean and the standard deviation of
their TVD, so that a correct sampler gives a ratio near one and a deviation $z$ of order one. This reference stays valid
when the output space is large against $S$, where the normal approximation of the shot noise overestimates it.

\paragraph{Oracles.} Each oracle shares no code with the sampler it checks. The strong simulation is SLOS~\cite{Heurtel2023SLOS},
in kernels of our code base: on the Fock input for single photons; on the coherent product state, layer by layer in the
photon number, for superposed sources, with loss as a thinning of the layers and the $x$-model as an enumeration over
the sources in the common mode; on the matrix-product state, traced or projected, for photonic MPS, with loss by
thinning; once per adaptive outcome for adaptive optics; and once per branch, summed coherently, for non-linear sites,
where it agrees to $10^{-16}$ with two dense references, the gate-by-gate state and the branch sum of permanents. Two
further checks use none of our code. For superposed sources, Perceval's SLOS backend~\cite{Perceval} on the coherent
product \texttt{StateVector} agrees within shot noise (Supplementary Material, Sec.~\ref{S-sec:perceval}), and the exact law
of the algorithm, enumerated over every source order and every branch, equals $q$ to $10^{-13}$ on six instances with
$n\le5$.

\noindent The Supplementary Material (Sec.~\ref{S-sec:valid}, Table~\ref{S-tab:valid}) lists $19$ instances: single photons, superposed sources with
loss and partial distinguishability, traced and projected MPS trains up to $L=3$ and $\chi=3$, lossy trains, adaptive
circuits with one to three levels, and one or two non-linear sites. On all of them the measured TVD stays within
$1.7$ standard deviations of an exact draw.

\subsection{Cost}
\label{sec:table}

\noindent Table~\ref{tab:cost} gives the cost of each sampler relative to the single-photon
sampler~\cite{CC2020} on an instance of the same size, both run on one NVIDIA H100 with Haar interferometers at $m=2n$;
what these factors imply for the size of the devices that can be simulated at a given rate is the subject of
Ref.~\cite{Frontier}. Each $n$ uses one Haar unitary, the QR decomposition of a complex Ginibre matrix with the phases of
the diagonal of $R$ fixed. The time of a sample is the wall-clock time of a sampling call, the GPU synchronised before and
after; the first call, which compiles the kernels, is excluded. Arithmetic is IEEE double precision (complex128).\footnote{\label{fn:exact}``Exact''
refers throughout to the ideal algorithms: in exact arithmetic their output law equals the target. Sec.~\ref{S-sec:division}
gives the measured rounding error of the cofactors, at rounding level at the sizes used here. Agreement of histograms with
strong simulation within shot noise is evidence for the implementation, not a proof of exact numerical sampling.}
The superposed-source costs are averages, over the binomial law of the photon number $k$, of the cost measured at each $k$:
the cost of a shot grows like $2^k$, and a plain mean over a short run would miss the rare large-$k$ shots.

\begin{table}[h]\centering\footnotesize
\setlength{\tabcolsep}{4pt}\begin{tabular}{L{2.9cm}L{3.7cm}L{1.7cm}L{5.5cm}}\toprule
sampler (Sec.) & instance & kind & cost, relative to the single-photon sampler of the same size\\\midrule
single photons~\cite{CC2020} & $n=30$ / $34$ photons & time & $1$\\
superposed sources (\ref{sec:sbs}) & $n=30$ sources, $\sin^2\alpha=1$ (offset empty) & time & $1.05$--$1.15$\\
 & $n=34$ sources, $\sin^2\alpha=1$ / $0.99$ / $0.95$ / $0.9$ / $0.8$ & time & $1.04\pm0.09$ / $0.97$ / $0.75$ / $0.52$ / $0.23$\\
photonic MPS (\ref{sec:ebs}) & linear cluster, $L=\chi=2$, $n=8$ / $12$ / $16$ & count & a step costs $R_j\chi^2$
single-photon steps; mean $R_j$ peaks at $7$ / $30$ / $96$ near $j\approx n/2.5$ and falls to $1$--$2$ on the last steps,
which carry the cost\\
adaptive (\ref{sec:adaptive}) & $n=30$--$38$, $K=1$ / $2$, Haar policy & count & $2.1$ / $4.1$ chains
per sample (mean)\\
 & $n=38$, $K=1$ / $2$ / $3$ & model & time $X=1.03$ / $1.07$ / $1.10$\\
non-linear site (\ref{sec:nlbs}) & one site, $\varphi=\pi/2$ ($B=2$) & count & acceptance $0.43$--$0.59$ against
$1/\|c\|_1^2=0.5$; branch permanent $\kappa=0.55\pm0.08$ samples\\
 & & model & $(1+\kappa B)\|c\|_1^2=2(1+2\kappa)\approx4.2$\\
 & $B=16$ & count &acceptance $0.054$--$0.077$ against $0.0625$\\\bottomrule
\end{tabular}
\caption{Cost of the samplers on one H100, $m=2n$, Haar interferometers, relative to the single-photon sampler on an
instance of the same size (for superposed sources: $n$ sources against $n$ photons); the single-photon sampler takes
$6.2$\,ms ($n=30$) and $69$\,ms ($n=34$) per sample. ``kind'' separates measured \underline{times},
measured \underline{counts} (term counts, chains, acceptances) and \underline{model} estimates. All samplers are exact (in the sense of
footnote~\ref{fn:exact}) and are validated in Sec.~\ref{sec:validation}.}\label{tab:cost}
\end{table}

\noindent The cost of superposed sources falls with the photon number at the exponent of single photons under loss
$\eta=\sin^2\alpha$, and the coherence enters the cost of a step through the offset: $s_0=n-k$ grows as the sources dim, and each Glynn term carries the factor
$|T_j|(d+1)/j$ of Table~\ref{tab:criterion}; at $\sin^2\alpha=1$ the offset is empty and the sampler runs the single-photon
step at $1.05$--$1.15$ times its cost ($n=30$; $1.04\pm0.09$ at $n=34$). Against single photons under loss
$\eta=\sin^2\alpha$ on the same $n=34$ instance the time ratio is $1.1\pm0.1$, $1.6\pm0.1$, $2.2\pm0.1$ and $4.1\pm0.1$
at $\sin^2\alpha=0.99$, $0.95$, $0.9$ and $0.8$. For photonic MPS the overhead is the term count $R_j$, which is small on the last steps that carry
the cost on the trains we measured ($n\le20$); how it grows with $n$ is open.
For adaptive optics the overhead is set by the photons read adaptively, $\bar r=nK/m$, not by the number of adaptive
modes, and only $K\propto m$ changes the exponent. A non-linear site costs $(1+\kappa B)\|c\|_1^2$, a constant for a
response of short period and polynomial for a generic one. For a fixed number of sites and adaptive modes, adaptive optics and
non-linear sites keep the exponential rate of the single-photon chain, superposed sources take that of single photons under
loss $\eta=\sin^2\alpha$, and for photonic MPS the rate depends on the growth of $R_j$.

\paragraph{Competing exact algorithms.} None of the exact samplers of Sec.~\ref{sec:intro} has a public implementation
specialised to these inputs; generic Fock-space simulators handle mid-circuit photon counting at the cost of strong
simulation. A cost-model comparison at the sizes of Table~\ref{tab:cost} (Supplementary Material, Sec.~\ref{S-sec:compete}) finds
every alternative slower by one to several orders of magnitude, conditional on its assumptions. For photonic MPS the
prefix-amplitude sampler~\cite{BGL2022} has a polynomial overhead over single photons, ours the term count $R_j$: if $R_j$
grows exponentially with $n$, the prefix-amplitude sampler is faster at large $n$ (Supplementary Material,
Sec.~\ref{S-sec:compete}).

\section{Conclusion}
\label{sec:conclusion}

Of the three properties behind Clifford and Clifford's sampler, the random order does not need the product input. The
set-chain lemma asks only for an identity between weights on input sets, and superposed sources and photonic MPS satisfy
it although neither is a product of photons; the hidden set of sources that supplied no photon, or of emissions traced
out, is carried by the chain and never has to be enumerated. The chain does need the orthonormality of the columns.
Where it fails --- on the non-adaptive modes of an adaptive circuit, between the branches of a non-linear site --- the
weights couple different hidden sets and the identity breaks; the samplers keep the chain for the part where it holds
and correct the rest by rejection, whose acceptance probability is known exactly.

A source in a coherent superposition of vacuum and one photon is sampled at the
exponential rate of a lossy single photon of the same brightness, and adaptive measurements on a fixed number of modes or a
non-linear phase of short period multiply the single-photon cost by a constant. None of these features, at that scale,
moves the exponential rate of exact simulation. For photonic MPS the question remains open. The cost of our sampler is set
by the term count $R_j$, small on the trains we measured, and whether it stays small as the trains grow decides whether
it or the prefix-amplitude sampler, whose overhead over single photons is polynomial, is the faster one on long trains.

Four questions remain open. The first is the growth of $R_j$ with $n$, which sets the rate for photonic MPS. The second is
whether a chain-consistent family exists for a coherent sum of linear branches with polynomially many terms per set; it
would make a weak non-linearity cheap where the rejection makes it costly, and we know neither such a family nor an
argument against one. The third is a lower bound on the cost of chain-consistent families for any of these inputs, since
the lemma says when a step is cheap but not that no cheaper sampler exists. The fourth concerns adaptive circuits whose
number of adaptive modes grows in proportion to the number of modes, where the rejection changes the exponent and we know
no sampler that avoids it.

The samplers are part of a larger study of the exact classical simulation of photonic devices~\cite{Frontier}. Their code
is open source (MIT licence) at \url{https://github.com/Quandela/photonic-engines} and on PyPI as
\texttt{photonic-engines}; the Lean proofs are available at \href{https://doi.org/10.5281/zenodo.24023537}{doi:10.5281/zenodo.24023537}.

\section*{Acknowledgements}
We thank Timothée Goubault de Brugière for useful conversations, and Rawad Mezher for his insightful feedback and
inspiration.
This work has been co-funded by the European Commission by the Horizon-CL4 program under the grant agreement 101135288 for the EPIQUE project, by the PROQCIMA and TUF-TOPIQC programs within the French National Quantum Strategy (France 2030).
The sampler code, the Lean proofs (Supplementary Material, Sec.~\ref{S-sec:lean}), and the
manuscript were written with the assistance of generative models from Anthropic (Claude). The algorithms, the
formalisation choices, the code architecture, the main definitions, the lemma and the propositions were designed and
written or reviewed by the authors, who take responsibility for the whole.


\begin{thebibliography}{99}
\bibitem{AA2013} S.~Aaronson and A.~Arkhipov, The computational complexity of linear optics, Theory Comput.\ 9, 143 (2013).
\bibitem{CC2018} P.~Clifford and R.~Clifford, The classical complexity of boson sampling, Proc.\ SODA 2018, doi:10.1137/1.9781611975031.10, arXiv:1706.01260.
\bibitem{CC2020} P.~Clifford and R.~Clifford, Faster classical boson sampling, Phys.\ Scr.\ 99, 065121 (2024), arXiv:2005.04214.
\bibitem{BrodOszmaniec} D.~J.~Brod and M.~Oszmaniec, Classical simulation of linear optics subject to nonuniform losses, Quantum 4, 267 (2020), doi:10.22331/q-2020-05-14-267, arXiv:1906.06696.
\bibitem{Renema2020SBS} J.~J.~Renema, Simulability of partially distinguishable superposition and Gaussian boson sampling, Phys.\ Rev.\ A 101, 063840 (2020), arXiv:1911.10112.
\bibitem{Renema2021marginals} J.~J.~Renema, Marginal probabilities in boson samplers with arbitrary input states, arXiv:2012.14917 (2020).
\bibitem{Schon2005} C.~Sch\"on, E.~Solano, F.~Verstraete, J.~I.~Cirac and M.~M.~Wolf, Sequential generation of entangled multiqubit states, Phys.\ Rev.\ Lett.\ 95, 110503 (2005).
\bibitem{Lindner2009} N.~H.~Lindner and T.~Rudolph, Proposal for pulsed on-demand sources of photonic cluster state strings, Phys.\ Rev.\ Lett.\ 103, 113602 (2009).
\bibitem{Huet2025} H.~Huet, P.~R.~Ramesh, S.~C.~Wein, N.~Coste, P.~Hilaire, N.~Somaschi, M.~Morassi, A.~Lema\^itre, I.~Sagnes, M.~F.~Doty, O.~Krebs, L.~Lanco, D.~A.~Fioretto and P.~Senellart, Deterministic and reconfigurable graph state generation with a single solid-state quantum emitter, Nat.\ Commun.\ 16, 4337 (2025), doi:10.1038/s41467-025-59693-3.
\bibitem{CMS2021} U.~Chabaud, D.~Markham and A.~Sohbi, Quantum machine learning with adaptive linear optics, Quantum 5, 496 (2021), arXiv:2102.04579.
\bibitem{OhLim2025adaptive} C.~Oh and Y.~Lim, Classical algorithms for measurement-adaptive Gaussian circuits, PRX Quantum 7, 010329 (2026), doi:10.1103/rpl7-ylg8, arXiv:2509.00746.
\bibitem{Spagnolo2021} N.~Spagnolo, D.~J.~Brod, E.~F.~Galv\~ao and F.~Sciarrino, Non-linear boson sampling, npj Quantum Inf.\ (2023), doi:10.1038/s41534-023-00676-x, arXiv:2110.13788.
\bibitem{Quesada2022} N.~Quesada, R.~S.~Chadwick, B.~A.~Bell, J.~M.~Arrazola, T.~Vincent, H.~Qi and R.~Garc\'ia-Patr\'on, Quadratic speed-up for simulating Gaussian boson sampling, PRX Quantum 3, 010306 (2022), arXiv:2010.15595.
\bibitem{QuesadaArrazola2020} N.~Quesada and J.~M.~Arrazola, Exact simulation of Gaussian boson sampling in polynomial space and exponential time, Phys.\ Rev.\ Research 2, 023005 (2020), arXiv:1908.08068.
\bibitem{DiasKoenig2025} B.~Dias and R.~K\"onig, On the sampling complexity of coherent superpositions, arXiv:2501.17071 (2025, v2 2026).
\bibitem{BGL2022} S.~Bravyi, D.~Gosset and Y.~Liu, How to simulate quantum measurement without computing marginals, Phys.\ Rev.\ Lett.\ 128, 220503 (2022), arXiv:2112.08499.
\bibitem{OhTreewidth2022} C.~Oh, Y.~Lim, B.~Fefferman and L.~Jiang, Classical simulation of boson sampling based on graph structure, Phys.\ Rev.\ Lett.\ 128, 190501 (2022), arXiv:2110.01564.
\bibitem{RahimiKeshari2016} S.~Rahimi-Keshari, T.~C.~Ralph and C.~M.~Caves, Sufficient conditions for efficient classical simulation of quantum optics, Phys.\ Rev.\ X 6, 021039 (2016), arXiv:1511.06526.
\bibitem{Bulmer2022} J.~F.~F.~Bulmer et al., The boundary for quantum advantage in Gaussian boson sampling, Sci.\ Adv.\ 8, eabl9236 (2022), arXiv:2108.01622.
\bibitem{Glynn2010} D.~G.~Glynn, The permanent of a square matrix, Eur.\ J.\ Combin.\ 31, 1887 (2010), doi:10.1016/j.ejc.2010.01.010.
\bibitem{IvanovGurvits2020} D.~A.~Ivanov and L.~Gurvits, Complexity of full counting statistics of free quantum particles in product states, Phys.\ Rev.\ A 101, 012303 (2020).
\bibitem{GPRS2019} R.~Garc\'ia-Patr\'on, J.~J.~Renema and V.~Shchesnovich, Simulating boson sampling in lossy architectures, Quantum 3, 169 (2019), doi:10.22331/q-2019-08-05-169, arXiv:1712.10037.
\bibitem{Heurtel2023SLOS} N.~Heurtel, S.~Mansfield, J.~Senellart and B.~Valiron, Strong simulation of linear optical processes, Comput.\ Phys.\ Commun.\ 291, 108848 (2023), arXiv:2206.10549.
\bibitem{Perceval} N.~Heurtel et al., Perceval: a software platform for discrete variable photonic quantum computing, Quantum 7, 931 (2023).
\bibitem{Frontier} J.~Senellart, B.~Stott, S.~Wein, E.~Cuenca Gomez, P.~Senellart, P.-E.~Emeriau and S.~Mansfield, The simulability frontiers of photonic quantum computing, in preparation (2026).
\bibitem{HowardCampbell2017} M.~Howard and E.~Campbell, Application of a resource theory for magic states to fault-tolerant quantum computing, Phys.\ Rev.\ Lett.\ 118, 090501 (2017), arXiv:1609.07488.
\bibitem{Bravyi2019} S.~Bravyi, D.~Browne, P.~Calpin, E.~Campbell, D.~Gosset and M.~Howard, Simulation of quantum circuits by low-rank stabilizer decompositions, Quantum 3, 181 (2019), arXiv:1808.00128.
\bibitem{Seddon2021} J.~R.~Seddon, B.~Regula, H.~Pashayan, Y.~Ouyang and E.~T.~Campbell, Quantifying quantum speedups: improved classical simulation from tighter magic monotones, PRX Quantum 2, 010345 (2021), arXiv:2002.06181.
\bibitem{Moylett2020} A.~E.~Moylett, R.~Garc\'ia-Patr\'on, J.~J.~Renema and P.~S.~Turner, Classically simulating near-term partially-distinguishable and lossy boson sampling, Quantum Sci.\ Technol.\ 5, 015001 (2020), arXiv:1907.00022.
\end{thebibliography}
\end{document}


\maketitle

\noindent Sections~\ref{sec:code}--\ref{sec:prior} give the pseudo-code, a cost comparison with the other exact
algorithms, the check against Perceval, the proof of the adaptive column draw and the list of related samplers.
Sections~\ref{sec:hardness}--\ref{sec:lean} treat the complexity of photonic matrix-product states (MPS), the failure of the chain
at a non-linear site, the validation table, loss and partial distinguishability, the leave-one-out values by division,
and the formal proofs in Lean. The notation is that of the main text; sections, tables, lemmas, propositions and remarks numbered without the
prefix S are those of the main text.

\section{Pseudo-code}
\label{sec:code}

\noindent We give the single-photon reference (Remark~2 of the main text) and the four samplers: superposed sources
(Sec.~3), photonic MPS (Sec.~4), adaptive linear optics (Sec.~5) and non-linear sites (Sec.~6). The boxes
show the lossless or lossy basic case; partial distinguishability is the extension of Sec.~\ref{sec:imperfections}, except
in Algorithm~\ref{alg:sbs}, which includes it.

\begin{algorithm}[H]
\caption{Reference: single photons, Clifford and Clifford with the collapse of repeated rows~\cite{CC2018,CC2020}.}\label{alg:fock}
\begin{algorithmic}[1]
\Require $U$ ($m\times m$), input modes $t_1..t_n$;\quad $A\gets U[:,t]$ ($m\times n$)
\State $\sigma\gets$ uniform order of the $n$ columns
\For{$k=1..n$}
  \State $T\gets\{\sigma_1..\sigma_k\}$;\quad $B\gets A[z_{<k},T]$ \Comment{$(k-1)\times k$, rows may repeat}
  \State $P_c\gets\per B_{\setminus c}$ ($B$ without column $c$) for all $c\in T$ \Comment{\Call{Cofactors}{B}}
  \State $\mathrm{amp}(r)\gets\sum_{c\in T}A_{rc}P_c$ for every output mode $r$;\quad draw $z_k\propto|\mathrm{amp}|^2$
\EndFor
\State \Return occupation of $(z_1..z_n)$
\Statex
\Function{Cofactors}{$B$} \Comment{one Glynn sum gives all $k$ cofactors}
  \State $(b_\nu,\mu_\nu)\gets$ distinct rows of $B$ and their multiplicities
  \State $v\gets0$;\quad $w\gets1$;\quad $s\gets\sum_\nu\mu_\nu b_\nu$;\quad $\mathrm{acc}_c\gets0$ for all $c$ \Comment{$v_\nu$: rows of group $\nu$ signed $-1$}
  \For{each $v\in\prod_\nu\{0..\mu_\nu\}$ in reflected mixed-radix Gray order} \Comment{$\prod_\nu(\mu_\nu+1)$ terms}
    \State on the step $v_\nu\to v_\nu\pm1$: $s\gets s\mp2b_\nu$;\quad $w\gets w\cdot$ (ratio of $(-1)^{v_\nu}\binom{\mu_\nu}{v_\nu}$)
    \State $\mathrm{acc}_c\gets\mathrm{acc}_c+w\prod_{j\ne c}s_j$ for all $c$ \Comment{all-but-one products}
  \EndFor
  \State \Return $\mathrm{acc}/2^{k-1}$ \Comment{full sum as written; pairing $v$ with $\mu-v$ halves it (as in~\cite{CC2018})}
\EndFunction
\end{algorithmic}
\end{algorithm}

\begin{algorithm}[H]
\caption{Superposed sources: random source order with offset start.}\label{alg:sbs}
\begin{algorithmic}[1]
\Require $U$ ($m\times m$), source modes $t_1..t_n$, $\alpha$; transmission $\eta$, common-mode probability $x$
\State draw the live sources (each with prob.\ $\cos^2\alpha+\eta\sin^2\alpha$), and among them those in the common
\Statex \hspace{\algorithmicindent}internal mode (each with prob.\ $x$);\quad no live source: \textbf{return} the vacuum
\State $\alpha'\gets\operatorname{atan2}(\sqrt\eta\sin\alpha,\cos\alpha)$ \Comment{$\tan\alpha'=\sqrt\eta\tan\alpha$, defined at $\alpha=\pi/2$}
\For{each live source $i$ in its own internal mode} \Comment{factorises out of the Dicke coherence}
  \State with prob.\ $\sin^2\alpha'$: one photon on output $r\sim|U_{r,t_i}|^2$ \Comment{otherwise vacuum}
\EndFor
\State $L\gets$ live sources in the common mode; draw $k\sim\mathrm{Bin}(|L|,\sin^2\alpha')$
\State $\sigma\gets$ uniform order of $L$;\quad $A\gets U[:,t_{\sigma}]$
\For{$j=1..k$}
  \State $T\gets$ first $|L|-k+j$ columns of $A$
  \State $P_x\gets\perrect A[z_{<j},T\setminus x]$ for all $x\in T$ \Comment{\Call{RectCofactors}{$A[z_{<j},T]$}}
  \State $w(r)\gets|\sum_{x\in T}A_{rx}P_x|^2$ for $r\in[m]$;\quad draw $z_j\propto w$
\EndFor
\State \Return occupation of $(z_1..z_k)$ plus the photons of the sources in their own mode
\Statex
\Function{RectCofactors}{$B$} \Comment{$B$: $q=j-1$ rows, $N=|T|$ columns, $s_0=N-1-q$}
  \State \textbf{if} $q=0$ \textbf{then return} $P\equiv1$
  \State $P\gets0$
  \State $(b_\nu,\mu_\nu)\gets$ distinct rows of $B$ and their multiplicities;\quad $a\gets\sum_\nu\mu_\nu b_\nu$;\quad $w\gets1$
  \For{each $v\in\prod_\nu\{0..\mu_\nu\}$ in reflected mixed-radix Gray order} \Comment{as in Algorithm~\ref{alg:fock}}
    \State on the step $v_\nu\to v_\nu\pm1$: $a\gets a\mp2b_\nu$;\quad $w\gets w\cdot$ (ratio of $(-1)^{v_\nu}\binom{\mu_\nu}{v_\nu}$)
    \State $P_x\gets P_x+w\,$\Call{Loo}{$a,x$} for all $x$
  \EndFor
  \State \Return $P/2^{q}$ \Comment{full sum as written; $e_q$ homogeneous, so pairing $v$ with $\mu-v$ halves it}
\EndFunction
\Statex
\Function{Loo}{$a,x$} \Comment{$e_q(a\setminus a_x)$, $d=\min(q,s_0)$}
  \If{$s_0=0$} \Return $\prod_{i\ne x}a_i$ \Comment{all-but-one product: the single-photon payload}
  \ElsIf{$s_0\le q$} \Return $[\mu^{s_0}]\,\mathrm{Pre}_x(\mu)\,\mathrm{Suf}_x(\mu)$ \Comment{$\prod_{i<x},\prod_{i>x}$ of $(\mu+a_i)$, degree $\le s_0$}
  \Else\ \Return $[\lambda^{q}]\,\mathrm{Pre}_x(\lambda)\,\mathrm{Suf}_x(\lambda)$ \Comment{$\prod_{i<x},\prod_{i>x}$ of $(1+\lambda a_i)$, degree $\le q$}
  \EndIf
\EndFunction
\end{algorithmic}
\end{algorithm}

\begin{algorithm}[H]
\caption{Photonic MPS: reduced-state chain with purified terms and matrix cofactors.}\label{alg:ebs}
\begin{algorithmic}[1]
\Require $U$, emission operators $M^{(s)}_l$ ($\chi\times\chi$) and input modes $t_{s,l}$ ($s=1..n$, $l=1..L$), $\ket{\mathrm{in}}$;
transmission $\eta$
\State draw the lost emissions (each with probability $1-\eta$); $\mathcal K\gets$ the kept ones, $k\gets|\mathcal K|$
\State draw the labels $x$ of the lost emissions and the final emitter state $\bra{\mathrm{out}}$ from their exact joint marginal
\Statex \Comment{label unravelling: each lost emission now carries one operator $M^{(s)}_{x_s}$; lossless: $\bra{\mathrm{out}}$ only}
\Statex \Comment{the kept state given these draws is normalised; its scale cancels from the row draws below}
\State $\sigma\gets$ uniform order of $\mathcal K$
\For{$j=1..k$}
  \State $T\gets\{\sigma_1..\sigma_j\}$ in emission order;\quad $\{(\ket{\mathrm{in}_e},\bra{\mathrm{out}_e},\tilde M_e)\}_{e\le R_j}\gets\Call{GapTerms}{T}$
  \For{each term $e$}
    \State $K^e_c\gets\Call{MatrixCofactors}{e,z_{<j},T}$ for $c\in T$
    \State $\mathrm{amp}_e(r)\gets\sum_{c\in T}\sum_lU_{r,t_{c,l}}\Tr\big(\tilde M^{(c)}_{e,l}K^e_c\big)$ for every output mode $r$
  \EndFor
  \State $w(r)\gets\sum_e|\mathrm{amp}_e(r)|^2$;\quad draw $z_j\propto w$ \Comment{$w(r)=f_j(z_{<j}r;T)$}
\EndFor
\State \Return occupation of $(z_1..z_k)$
\Statex
\Function{GapTerms}{$T$} \Comment{purification of $\rho_T$: $\rho_T=\sum_e\ket{\psi_e}\!\bra{\psi_e}$}
  \For{each gap (maximal run of emissions $a+1..b-1$ not in $T$, prefix and suffix runs included)}
    \State Kraus operators $\{Q_i\}$ of $\rho\mapsto\sum_xA_x\rho A_x^\dagger$, $A_x=M^{(b-1)}_{x_{b-1}}\cdots M^{(a+1)}_{x_{a+1}}$
    \Statex \hspace{\algorithmicindent}lost emissions inside the run enter $A_x$ with their drawn operator, not summed
    \Statex \hspace{\algorithmicindent}by SVD of the matrix with columns $\operatorname{vec}(A_x)$ (the Choi matrix), reduced emission by emission
    \Statex \hspace{\algorithmicindent}(rank $\le\chi^2$; $\le\chi$ at the ends; Prop.~2)
  \EndFor
  \State one term per choice of $i$ in every gap: $\tilde M^{(a)}_l\gets Q_iM^{(a)}_l$ for the kept emission $a$ before the gap,
  \Statex \hspace{\algorithmicindent}prefix gap into $\ket{\mathrm{in}_e}$, suffix gap into $\bra{\mathrm{out}_e}$; drop zero-norm chains
  \State \Return the $R_j$ terms
\EndFunction
\Statex
\Function{MatrixCofactors}{$e,z_{<j},T$} \Comment{Glynn over the $j-1$ placed rows, $\chi\times\chi$ products}
  \State \textbf{if} $j=1$ \textbf{then return} $K_c=\ket{\mathrm{in}_e}\!\bra{\mathrm{out}_e}$ for the single $c\in T$ \Comment{no placed row}
  \State $K_c\gets0$ for $c\in T$
  \For{$\delta\in\{+1\}\times\{\pm1\}^{j-2}$ in Gray order} \Comment{2020 collapse on repeated rows: mixed-radix walk}
    \State $s_{c,l}\gets\sum_i\delta_iU_{z_i,t_{c,l}}$;\quad $G_c\gets\sum_ls_{c,l}\tilde M^{(c)}_{e,l}$
    \State $p_c\gets G_{c-1}\cdots G_1\ket{\mathrm{in}_e}$,\quad $q_c^T\gets\bra{\mathrm{out}_e}G_{j}\cdots G_{c+1}$ \Comment{prefix / suffix sweeps}
    \State $K_c\gets K_c+(\prod_i\delta_i)\,p_c\,q_c^T$
  \EndFor
  \State \Return $2^{-(j-2)}K_c$
\EndFunction
\end{algorithmic}
\end{algorithm}

\begin{algorithm}[H]
\caption{Adaptive linear optics: exact samples $\omega=(p,s)$.}\label{alg:adaptive}
\begin{algorithmic}[1]
\Require $U_0$ ($m\times m$), input modes $t$ ($n$ photons), levels $A_1..A_\Lambda$ (adaptive modes $a_1..a_K$ in measurement
order), policy $V_\lambda(p_{\le\lambda})$
\State $p\gets()$,\quad $U\gets U_0$
\For{$\lambda=1..\Lambda$} \Comment{stage 1: the adaptive outcome, level by level}
  \If{the $K_\lambda$-mode graded table fits}
    \State $M\gets\Call{BinomialMoments}{U,t,a_1..a_{K_\lambda}}$;\quad $P\gets$ binomial inverse of $M$
    \State draw the level-$\lambda$ counts from $P[p_{<\lambda},\cdot\,]/\sum P[p_{<\lambda},\cdot\,]$
  \Else \Comment{beyond the tables: a level chain}
    \Repeat\ $\rho\gets\Call{PrefixChain}{U,t,a^{p_{<\lambda}},\{a_1..a_{K_{\lambda-1}}\}}$ \Until{$\rho\neq$ \textsc{abort}}
    \State level-$\lambda$ counts $\gets$ counts of $\rho$ on $A_\lambda$
  \EndIf
  \State $p\gets p\oplus$ (level-$\lambda$ counts);\quad $U\gets E(V_\lambda(p))\,U$ \Comment{feed-forward on the unmeasured modes}
\EndFor
\State $a^p\gets(a_1^{\times p_1},\dots,a_K^{\times p_K})$,\quad $r\gets|p|$ \Comment{stage 2: the final counts given $p$}
\Repeat\ $\rho\gets\Call{PrefixChain}{U,t,a^p,\{a_1..a_K\}}$ \Until{$\rho\neq$ \textsc{abort}} \Comment{acceptance $P(p)/M[p]$}
\State $s\gets$ counts of $\rho_{r+1..n}$ on the final modes;\quad \Return $(p,s)$
\Statex
\Function{BinomialMoments}{$U,t,\mathcal A$} \Comment{Ivanov and Gurvits~\cite{IvanovGurvits2020}, as used in~\cite{OhLim2025adaptive}}
  \State $V\gets U[\mathcal A,t]$;\quad $G\gets1$ \Comment{polynomial in $x_1..x_{|\mathcal A|},w_1..w_{|\mathcal A|}$, graded by total degree}
  \For{$i=1..n$} $G\gets G\cdot\big(1+(\textstyle\sum_l\bar V_{li}x_l)(\sum_lV_{li}w_l)\big)$ \EndFor
  \State \Return $M[c]=c!\,[x^cw^c]\,G$ \Comment{$=\E[\prod_l\binom{N_l}{c_l}]$}
\EndFunction
\end{algorithmic}
\end{algorithm}

\begin{algorithm}[H]
\caption{The chain started from a forced prefix (Lemma~1(c) of the main text), with early abort.}\label{alg:prefix}
\begin{algorithmic}[1]
\Function{PrefixChain}{$U,t,\text{prefix }(r\text{ rows}),\text{avoid}$}
  \State $S\gets\Call{PrefixColumns}{U,t,\text{prefix}}$ \Comment{$\Pr[S]\propto|\per U[\text{prefix},S]|^2$}
  \State column order $\gets S$, then the other $n-r$ columns in uniform random order;\quad $\rho_{1..r}\gets$ prefix
  \For{$k=r+1..n$} \Comment{ordinary Clifford--Clifford steps}
    \State $w_j\gets|\per U[(\rho_{<k},j),\text{columns}_{\le k}]|^2$ for every output mode $j$ \Comment{Laplace, one Glynn sum}
    \State draw $\rho_k\propto w$; \textbf{if} $\rho_k\in$ avoid \textbf{then return} \textsc{abort}
  \EndFor
  \State \Return $\rho$
\EndFunction
\Statex
\Function{PrefixColumns}{$U,t,\text{prefix}$} \Comment{polynomial: the prefix has $K$ distinct rows}
  \State $B_j\gets U[a,t_j]$ for $j=1..n$ \Comment{$K$-vectors}
  \State $F_{n+1}\gets1$;\quad \textbf{for} $j=n..1$: $F_j\gets F_{j+1}\cdot\big(1+(B_j\cdot x)(\bar B_j\cdot w)\big)$, truncated at degree $p$
  \State $g\gets1$;\quad $S\gets\emptyset$ \Comment{$g$: polynomial in $x$, degree $\le p$}
  \For{$j=1..n$}
    \State $g_{\rm in}\gets g\cdot(B_j\cdot x)$
    \State $w_{\rm in}\gets\sum_{a,b}g_{\rm in}[a]\,\overline{g_{\rm in}[b]}\,F_{j+1}[p-a,p-b]$;\quad $w_{\rm out}\gets$ the same with $g$
    \State with probability $w_{\rm in}/(w_{\rm in}+w_{\rm out})$: $S\gets S\cup\{j\}$, $g\gets g_{\rm in}$
  \EndFor
  \State \Return $S$ \Comment{$|S|=r$}
\EndFunction
\end{algorithmic}
\end{algorithm}

\begin{algorithm}[H]
\caption{Non-linear sites: rejection from the incoherent branch mixture.}\label{alg:nlbs}
\begin{algorithmic}[1]
\Require interferometers $U_0..U_L$, sites $(\ell,x,f)$ (response $e^{-if(\hat n_x)}$ between $U_{\ell-1}$ and $U_\ell$), input
modes $t$; transmissions $\eta$ (before the sites), $\eta_{\rm post}$ (after)
\State draw the surviving photons (each with probability $\eta$); $n\gets$ their number, $t\gets$ their modes
\State \textbf{if} $n=0$ \textbf{then return} the vacuum
\State for each site: $(\theta_\ell,c_\ell)\gets\Call{Branches}{f,n}$ \Comment{the expansion of this stream's own photon count}
\State $U_b\gets U_LD_L(\theta_b)\cdots D_1(\theta_b)U_0$,\quad $c_b\gets\prod_{\rm sites}c_{\ell(\rm site)}$ for every branch $b$
\Repeat
  \State draw $b$ with probability $|c_b|/\|c\|_1$;\quad $s\gets$ one Fock sample of $U_b$ on $t$ \Comment{Clifford--Clifford, 2020 kernels}
  \State $a_{b'}\gets\per U_{b'}[s,t]$ for every branch $b'$ \Comment{batched Laplace cofactors}
  \State accept with probability $\big|\sum_{b'}c_{b'}a_{b'}\big|^2\big/\big(\|c\|_1\sum_{b'}|c_{b'}||a_{b'}|^2\big)$
\Until{accepted} \Comment{acceptance exactly $1/\|c\|_1^2$}
\State thin $s$ binomially with $\eta_{\rm post}$;\quad \Return $s$
\Statex
\Function{Branches}{$f,n$} \Comment{$e^{-if(\hat n_x)}=\sum_\ell c_\ell e^{i\theta_\ell\hat n_x}$ exactly on $\le n$ photons}
  \State $g(j)\gets e^{-if(j)}$ for $j=0..n$
  \State $q\gets$ smallest $q\le n$ with $g(j+q)/g(j)$ constant $=e^{iaq}$ on $j=0..n-q$ \Comment{$q=n$ always works}
  \State $h(j)\gets g(j)e^{-iaj}$ ($q$-periodic);\quad $c_\ell\gets\frac1q\sum_{j<q}h(j)e^{-2\pi i\ell j/q}$,\quad $\theta_\ell\gets a+2\pi\ell/q$
  \State \Return $\{(\theta_\ell,c_\ell):c_\ell\neq0\}$ \Comment{every non-zero coefficient: the expansion is exact}
\EndFunction
\end{algorithmic}
\end{algorithm}

\noindent\emph{Implementation note.} In floating point the transform returns coefficients at rounding level where the exact
ones vanish (e.g.\ at $\varphi=\pi/2$). Our code drops those with $|c_\ell|<10^{-14}$ and checks that the remaining
expansion reproduces $e^{-if(j)}$ on $j=0..n$ to $\varepsilon=10^{-12}$. The site is then replaced by a diagonal operator
within $\varepsilon$ of it on every Fock state of at most $n$ photons in mode $x$, so the output state moves by at most
$\varepsilon$ per site in norm and the output law by at most $2\varepsilon$ per site in total variation.

\section{Comparison with other exact algorithms}
\label{sec:compete}

Each sampler is compared with the other exact samplers that apply to its input (Sec.~1 of the main text and
Table~\ref{tab:prior}). We did not benchmark a public implementation of these alternatives;
Table~\ref{tab:compete} compares cost models, in units of single-photon samples (or steps) of the same size, the unit of
each cell being stated in it. Assumptions: a Haar interferometer on $m=2n$ modes decomposed into $g\approx m^2/2=2n^2$
beam splitters for the prefix-amplitude sampler of Ref.~\cite{BGL2022}, with $O(1)$ amplitudes per beam splitter, each a
permanent of the circuit truncated after that beam splitter and costing half a single-photon sample (the last chain step),
times $\chi^2$ for a matrix-valued Glynn sum. This is our adaptation of the gate-by-gate construction of
Ref.~\cite{BGL2022} to linear optics, not a cost stated there. A branch permanent costs $\kappa=0.55$ single-photon
samples (measured).

\begin{table}[H]\centering\footnotesize
\setlength{\tabcolsep}{4pt}\begin{tabular}{L{3.0cm}L{3.9cm}L{3.3cm}L{3.6cm}}\toprule
sampler, size & best prior exact alternative & its cost (estimate) & ours\\\midrule
superposed sources & generic chain~\cite{Renema2021marginals}; component rejection~\cite{DiasKoenig2025};
prefix amplitudes~\cite{BGL2022} & $\binom nk^2$ pairs per step; acceptance $\binom nk^{-1}$ per sector; $\sim2n^2$ amplitudes
of cost $2^kk\,n$ & $2^kn^2$ per sample (the exponent of lossy single photons)\\
photonic MPS, $n=20$ & prefix amplitudes~\cite{BGL2022} & $\approx4n^2\approx1.6\cdot10^3$ per step & $R_j\chi^2\approx4$--$8$
per step where the cost is (measured $R_j=1$--$2$)\\
adaptive, $K\le3$, $n=38$ & prefix amplitudes~\cite{BGL2022}; Gaussian model only~\cite{OhLim2025adaptive} & $\sim2n^2$ permanents
& $2.0$--$6.8$ chains started (cost model), mostly aborted early: time ratio $X=1.03$--$1.10$\\
non-linear, $\varphi=\pi/2$, $n=36$ & gadget~\cite{Spagnolo2021}; prefix amplitudes~\cite{BGL2022} & gadget: $\sim\varrho^{n}\approx2\cdot10^9$
before the herald rate; $2n^2B\kappa\approx3\cdot10^3$ & $(1+\kappa B)\|c\|_1^2=4.2$\\
non-linear, generic $\varphi$, $n=40$ & component rejection~\cite{DiasKoenig2025}; prefix amplitudes~\cite{BGL2022} &
$B(1+\kappa B)\|c\|_2^2\approx9.2\cdot10^2$; $\approx7\cdot10^4$ & $(1+40\kappa)\,33.2\approx7.6\cdot10^2$\\\bottomrule
\end{tabular}
\caption{Cost-model comparison with the exact alternatives, in single-photon-sample units at the stated sizes (assumptions in
the text). The ``ours'' column is measured on our implementation, except the MPS row (the step cost of Table~1 of the
main text with the measured term count) and the adaptive row (the cost model of Sec.~5 of the main text); the alternatives are cost models, not timed. Under these assumptions no
alternative is cheaper at these sizes; the MPS row may cross over at large $n$ (text).}\label{tab:compete}
\end{table}

\noindent Under the assumptions above, every modelled alternative is slower at these sizes by one to several orders of
magnitude. The comparison is conditional on them: none of the alternatives was timed on our machine, and a better
implementation of one could narrow the gaps.

\paragraph{Photonic MPS at large $n$.} At large $n$ the prefix-amplitude sampler may become the faster one. It works
with amplitudes of the pure photonic state and never forms the reduced states of Sec.~4 of the main text, so its overhead over single
photons is polynomial ($\sim4n^2$ single-photon steps), while ours is the term count, a step costing about $R_j\chi^2$
single-photon steps. How $R_j$ grows with $n$ is not known. On the trains we measured ($n\le20$) it is $1$--$2$ on the
last steps, which carry the cost, and our gapwise decomposition is within $1.02$--$1.11\times$ of the rank of $\rho_T$
weighted by step cost; but the bound of Proposition~2 grows exponentially with the number of gaps. If $R_j$ stays small on
the costly steps, our overhead is a constant factor; if it grows exponentially with $n$, the prefix-amplitude sampler
is faster once $R_j\chi^2$ exceeds $\sim4n^2$. A cost model of $R_j$ would locate the crossover, and a sampler that uses the
term expansion where $R_j$ is small and pure-state amplitudes where it is large would combine the two.

\section{Superposed sources against Perceval}
\label{sec:perceval}

\begin{table}[H]\centering\small
\begin{tabular}{cccc ccc}\toprule
$n$ & $m$ & $\alpha$ & $\eta$ & TVD to Perceval & shot noise & coherent vs incoherent\\\midrule
4 & 8 & $\pi/4$ & 1 & 0.0257 & 0.0263 & 0.374\\
6 & 12 & $\pi/4$ & 1 & 0.0991 & 0.1002 & 0.394\\
5 & 6 & 1.1 & 1 & 0.0365 & 0.0362 & 0.193\\
7 & 7 & $\pi/4$ & 1 & 0.0586 & 0.0580 & 0.421\\
5 & 8 & $\pi/4$ & 0.7 & 0.0281 & 0.0275 & 0.268\\\bottomrule
\end{tabular}
\caption{The superposed-source sampler (Sec.~3 of the main text) against Perceval's SLOS backend~\cite{Perceval} on the
coherent product \texttt{StateVector}, an oracle outside our code base; lossy points as the enumerated mixture over live
sources. $4\cdot10^4$ samples per point (\texttt{bench/sbs\_chain\_rso.py}, CPU). Shot noise: TVD of an exact multinomial
draw of the same size. Last column: TVD between the coherent law and the incoherent mixture of the same sources.}\label{tab:sbs-perceval}
\end{table}

\noindent The same test separates the coherent law from the incoherent mixture by a TVD of $0.19$--$0.42$ (last column).
Beyond strong-simulation reach ($n=14$, $m=28$) the mean occupation of every output mode matches the coherent prediction
$\sum_i|U_{ji}|^2(p-\gamma^2)+|\sum_iU_{ji}\gamma|^2$, with brightness $p=\sin^2\alpha$ and coherence
$\gamma=\cos\alpha\sin\alpha$, to $|z|\le1.8$, and misses the incoherent one by $|z|=22.1$ (one random instance,
$3000$ samples); this first-moment check
separates coherent from incoherent sources but cannot validate the full law.

\section{The column draw of the adaptive sampler}
\label{sec:columns}

Let the forced prefix have $K$ distinct rows $a_1..a_K$ with multiplicities $p$, $|p|=r$, and $B_j=U^p[a,t_j]\in\C^K$
(Sec.~5 of the main text). For $|S|=r$, a permanent with repeated rows is a coefficient,
$\per U^p[a^p,S]=p!\,[x^p]\prod_{j\in S}(B_j\cdot x)$ with $p!=\prod_lp_l!$: expanding the product, the coefficient of
$x^p$ sums over the assignments of the columns of $S$ to the labels $l$ with counts $p$, and each assignment stands for
$\prod_lp_l!$ permutations. Hence $|\per U^p[a^p,S]|^2=(p!)^2\,[x^pw^p]\prod_{j\in S}(B_j\cdot x)(\bar B_j\cdot w)$.
Deciding the columns in the order $1..n$, with $S_<$ the columns already taken and $g=\prod_{j\in S_<}(B_j\cdot x)$, the
total weight of the completions is
\begin{gather*}
  \sum_{S'\subseteq\{j..n\}}[x^pw^p]\,g(x)\bar g(w)\prod_{i\in S'}(B_i\cdot x)(\bar B_i\cdot w)
  =[x^pw^p]\,g(x)\bar g(w)F_j(x,w),\\
  F_j=\prod_{i\ge j}\big(1+(B_i\cdot x)(\bar B_i\cdot w)\big),
\end{gather*}
where truncation at degree $p$ keeps exactly the completions with $|S|=r$. Taking column $j$ with probability
$w_{\rm in}/(w_{\rm in}+w_{\rm out})$, the two completion weights with and without it, therefore draws $S$ with probability
$\propto|\per U^p[a^p,S]|^2$, the posterior of Lemma~1(c) of the main text. The tables have $\prod_l(p_l+1)^2$ entries,
polynomial for fixed $K$ (Algorithm~\ref{alg:prefix}).

\section{Prior exact samplers}
\label{sec:prior}

\begin{table}[H]\centering\footnotesize
\setlength{\tabcolsep}{4pt}\begin{tabular}{L{4.1cm}L{4.4cm}L{5.6cm}}\toprule
method & applies to & cost structure, and relation to the main text\\\midrule
chain over the true marginal~\cite{Renema2021marginals} & any input state & pairs of input components $\times$ hidden
subsets per step: up to $\binom nk^2$ for superposed sources (direct expansion, late steps), up to $L^{2n}$ for an MPS
train (Remark~1)\\
component rejection~\cite{DiasKoenig2025} & coherent superposition of $N_{\rm c}$ components & $N_{\rm c}\|\beta\|_2^2$ proposals;
$N_{\rm c}=2^n$ for $n$ superposed sources in the Fock basis; its principle, reweighted, in Sec.~6\\
prefix amplitudes~\cite{BGL2022} & any circuit & $O(g)$ amplitudes per sample, no marginals; Sec.~\ref{sec:compete}\\
repeated input modes~\cite{BrodOszmaniec} & Fock input with repeated modes & Clifford--Clifford extended\\
Gaussian chains~\cite{QuesadaArrazola2020,Quesada2022,Bulmer2022} & Gaussian inputs & loop hafnians
over output modes, $2^{N/2}$ in the detected photons; out of scope\\
bounded treewidth~\cite{OhTreewidth2022}; phase space~\cite{RahimiKeshari2016} & structured interferometers; classical or
noisy light & tractable marginals; efficient\\
adaptive Gaussian~\cite{OhLim2025adaptive} & Gaussian input and measurement, $O(\log m)$ adaptive modes & efficient
exact sampler; Fock input with PNR final detection in Sec.~5\\
adaptive linear optics~\cite{CMS2021} & Fock input, adaptive measurements & estimates output probabilities and
overlaps; not a sampler of the full output\\
noisy adaptive circuits~\cite{GoubaultHeurtel2025} & Fock input, feed-forward, noise & strong simulation: output
probabilities, amplitudes as derivatives of a polynomial; not a sampler\\
Kerr gadget~\cite{Spagnolo2021} & one Kerr site & linear optics with ancillas and postselection; Sec.~6 needs no
ancilla\\\bottomrule
\end{tabular}
\caption{Exact samplers related to the inputs and circuits of the main text (sections and remarks refer to the main
text).}\label{tab:prior}
\end{table}

\section{Complexity of photonic matrix-product states}
\label{sec:hardness}

The model of Sec.~4 of the main text contains single-photon boson sampling: with $\chi=1$, $L=1$ and $M^{(s)}_1=1$ the
train is $n$ single photons in distinct modes, so exact sampling from photonic MPS is at least as hard, in the
worst case, as from Fock inputs~\cite{AA2013}. For a \emph{fixed} train, Ref.~\cite{Estimator} (Supplementary
Material, Proposition S2) proves more
for the projected register (a fixed $\bra{\mathrm{out}}$, $\ket\psi$ pure), assuming the tensors $M^{(s)}_l$, $\ket{\mathrm{in}}$, $\bra{\mathrm{out}}$
are polynomially bounded and given exactly and a label word $x^0$ with $\psi(x^0)\ne0$ is known. For $m\ge(L+1)n$,
computing the amplitude $\sum_x\psi(x)\per U[z,t_x]$ of a collision-free output $z$ exactly is $\#$P-hard, and so is
approximating its squared modulus to within a multiplicative factor $g\in[1,\mathrm{poly}(n)]$, by the corresponding
result for $|\per A|^2$~\cite{AA2013}: the interferometer is
chosen to carry a scaled copy of an arbitrary matrix on the columns $t_{x^0}$ and zeros on the $(L-1)n$ other label
columns, so that only $x^0$ contributes. Computing the polynomial $F_\psi(X)=\sum_x\psi(x)\per X[\cdot,t_x]$ exactly with
probability $\ge3/4+1/\mathrm{poly}(n)$ over standard complex Gaussian $X\in\C^{n\times Ln}$ is as hard: such an algorithm
$\mathcal O$ gives $\mathrm P^{\#\mathrm P}\subseteq\mathrm{BPP}^{\mathrm{NP}^{\mathcal O}}$. Hardness of
\emph{approximate} sampling in the sense of Ref.~\cite{AA2013} is only conditional. It needs a multiplicative-estimation
conjecture and an anticoncentration conjecture for $|F_\psi(X)|^2$, neither known to follow from the Fock case, and a
hiding step that padding to an $Ln\times Ln$ block takes from Ref.~\cite{AA2013} only for $m$ of $(Ln)^5$ type. For
single photons, Ref.~\cite{Bouland2026} gives hardness evidence with a linear number of modes; it does not by itself
establish the analogous average-case statement for a fixed non-trivial photonic MPS.

The worst-case statements extend to the traced register. With the other label columns set to zero, a collision-free output
of the traced train has probability $w\,|\per A'|^2$, $A'$ the scaled copy, where
$w=\|M^{(n)}_{x^0_n}\cdots M^{(1)}_{x^0_1}\ket{\mathrm{in}}\|^2$ is positive and computable in polynomial time for a
suitable known $x^0$; exact evaluation and multiplicative approximation of its output probabilities are therefore as hard as
for the projected register. We distinguish three questions: exact evaluation of probabilities (hard in the worst case, for
both registers), exact sampling (at least as hard as single photons, by the embedding $\chi=L=1$), and approximate
sampling in total variation for a fixed non-trivial train, which needs the average-case conjectures above and is open for
the traced register, whose output law mixes the final emitter states, and for the lost labels of a lossy train. These
statements do not bear on the cost measurements of the main text: the $m\ge(L+1)n$ requirement belongs to this
construction, and $m=2n=Ln$ in the caterpillar tests, which falls outside it, is not evidence of easiness either.

\section{Why the random-order chain fails at a non-linear site}
\label{sec:nlbs-fail}

In the notation of Sec.~6 of the main text, the marginal of the first $k$ ordered rows of the circuit $\sum_bc_bU_b$ is
\begin{gather*}
  q_k(z)=\tfrac{(n-k)!}{n!}\sum_{b,b'}c_b\bar c_{b'}\sum_{|S|=|T|=k}\per U_b[z,t_S]\,\overline{\per U_{b'}[z,t_T]}\;
  \per G_{bb'}[t_{T^c},t_{S^c}],\qquad G_{bb'}=U_{b'}^\dagger U_b,\\
  \textstyle\sum_{y\in[m]^{n-k}}\per U_b[y,t_{S^c}]\,\overline{\per U_{b'}[y,t_{T^c}]}=(n-k)!\,\per G_{bb'}[t_{T^c},t_{S^c}];
\end{gather*}
the rows of $\per G_{bb'}[t_{T^c},t_{S^c}]$ are the input modes of the photons $T^c$ left by the conjugated amplitude and
its columns those of the $S^c$ left by the other (second line). For a linear circuit $G=1$ and the last factor is $\delta_{ST}$,
which is Remark~2 of the main text. For a single site, $U_b=U_1D_x(\theta_b)U_0$ and
$G_{bb'}=1+\big(e^{i(\theta_b-\theta_{b'})}-1\big)\,ww^\dagger$ with $w=U_0^\dagger e_x$: the rank-one term couples
$S\neq T$. With several sites $G_{bb'}-1$ need not have rank one (two $\pi$ phases on different modes with identity
interferometers and one branch pair give $G=-1$ on those two modes); the general form above still holds. The true-marginal chain exists, but evaluated as written its quadratic form has $B^2\binom nj^2$ pairs of
branches and sets per candidate at step $j$, each with a permanent of size $n-j$, up to $4^n$ pairs at $j\approx n/2$; we
know no rewriting that reuses structure across the pairs as Glynn's sum does for one set. We found no set-indexed
purification of the kernel $c_b\bar c_{b'}\per G_{bb'}[t_{T^c},t_{S^c}]$ with few terms per set: purifying the traced photons in the
basis after the site gives a sum of squares indexed by their $m^{n-k}$ configurations, and splitting each traced photon into
``on $x$'' and ``not on $x$'' reintroduces the coupling through the rank-$(m-1)$ projector.

\section{Validation table}
\label{sec:valid}

\begin{table}[H]\centering\footnotesize
\setlength{\tabcolsep}{4pt}\begin{tabular}{L{2.1cm}L{5.4cm}ccccc}\toprule
sampler & instance & $n$ & $m$ & TVD & exact draw & $z$\\\midrule
single photons & Haar & 5 & 10 & 0.0722 & 0.0739 & $-1.0$\\
 & Haar & 6 & 12 & 0.1892 & 0.1896 & $-0.3$\\
superposed & $\alpha=\pi/4$ & 6 & 12 & 0.1031 & 0.1036 & $-0.3$\\
sources & $\alpha=1.1$ & 5 & 6 & 0.0376 & 0.0355 & $+1.5$\\
 & $\alpha=\pi/4$, $\eta=0.7$ & 6 & 12 & 0.0657 & 0.0650 & $+0.6$\\
 & $\alpha=\pi/4$, $x=0.8$ & 6 & 12 & 0.1152 & 0.1139 & $+0.9$\\
 & $\alpha=\pi/4$, $\eta=0.8$, $x=0.8$ & 6 & 12 & 0.0839 & 0.0849 & $-0.6$\\
photonic & linear cluster, traced (state of the cost table) & 4 & 8 & 0.0342 & 0.0332 & $+0.7$\\
 & caterpillar, projected & 5 & 10 & 0.0810 & 0.0797 & $+0.9$\\
MPS & random, $L=3$, $\chi=2$, traced$^\dagger$ & 3 & 9 & 0.0212 & 0.0223 & $-0.8$\\
 & random, $L=2$, $\chi=3$, traced$^\dagger$ & 3 & 7 & 0.0177 & 0.0166 & $+0.7$\\
 & linear cluster, traced, $\eta=0.7$ & 4 & 8 & 0.0392 & 0.0382 & $+0.8$\\
adaptive & $K=3$, two levels & 3 & 6 & 0.0132 & 0.0125 & $+0.4$\\
 & $K=2$, one level & 5 & 10 & 0.0752 & 0.0750 & $+0.1$\\
 & $K=3$, three levels & 4 & 8 & 0.0317 & 0.0309 & $+0.6$\\
non-linear & one Kerr site, $\varphi=\pi/2$ & 3 & 6 & 0.0131 & 0.0123 & $+0.6$\\
site & one Kerr site, $\varphi=\pi/4$ & 3 & 6 & 0.0121 & 0.0121 & $+0.1$\\
 & one Kerr site, $\varphi=0.9$ & 3 & 6 & 0.0141 & 0.0133 & $+0.5$\\
 & two sites in sequence & 3 & 6 & 0.0165 & 0.0136 & $+1.7$\\\bottomrule
\end{tabular}
\caption{Validation against strong simulation on one H100, through the calls of the cost measurements (Sec.~8 of the main
text), $S=4\cdot10^4$ samples per instance, Haar interferometers. ``Exact draw'': mean TVD of $40$ exact multinomial draws
of size $S$ from the exact law; $z$: the deviation of the measured TVD from it in standard deviations of those draws
(\texttt{bench/validate\_methods.py}). $^\dagger$Outside the range of the GPU kernels of this sampler ($\chi=2$, $L\le2$): the
reference implementation of the same algorithm, on the same GPU.}\label{tab:valid}
\end{table}

\section{Loss and partial distinguishability}
\label{sec:imperfections}

Loss is uniform, one transmission $\eta$ per photon at the input; partial distinguishability is the mixture ($x$-)model,
each photon or source independently in a common internal mode with probability $x$ and otherwise in an internal mode of
its own, orthogonal to all others, so two photons overlap with probability $x^2$. A general Gram matrix of internal
states is not covered by any of the samplers. Except for superposed sources, the extensions below are not part of the
algorithm boxes of Sec.~\ref{sec:code}, which show the basic cases.

\paragraph{Single photons.} Loss draws the surviving photons ($k\sim\mathrm{Bin}(n,\eta)$) and the chain runs on their
columns; distinguishability draws which survivors share the common mode, the others being routed classically by
$|U_{\cdot t_i}|^2$. The cost is that of the interfering photons.

\paragraph{Superposed sources.} Loss $\eta$ maps each source to a mixture of vacuum and a source at a smaller $\alpha'$,
$\tan\alpha'=\sqrt\eta\tan\alpha$, drawn per source before the chain. In the $x$-model, a source in its own mode factorises
out of the Dicke coherence: its photon is present with probability $\sin^2\alpha'$ and is routed by $|U_{\cdot t_i}|^2$, and
the chain runs on the live sources in the common mode. Both mixtures are drawn exactly (Algorithm~\ref{alg:sbs}).
Non-uniform loss, $\eta_i$ per source, gives live sources with different amplitudes, $\tan\alpha_i'=\sqrt{\eta_i}\tan\alpha$,
which the unequal-source chain of Sec.~3 of the main text covers.

\paragraph{Photonic MPS: unravelling the trace.} Two kinds of emission lie outside the active set at step $j$:
\emph{lost} ones, traced to the end, and \emph{kept} ones not yet placed. A trace is a label measurement whose outcome is
forgotten, so the labels of the lost emissions (and the final emitter state) can be drawn first, from their exact joint
marginal. Each lost emission then carries one fixed operator $M^{(s)}_{x_s}$, and the kept emissions are in a pure state, a
matrix-product state of bond dimension $\chi$ on $k$ emissions; the chain runs on it as on a lossless train of $k$
emissions. At step $j$ the gaps of Proposition~2 of the main text are made of kept, not yet placed emissions only (the lost
ones enter as fixed factors), so $R_j$ is that of the lossless $k$-emission train, and the last step has $R_k=1$. Without
this draw the lost emissions widen every gap and remain traced to the end; an inner gap of lost emissions contributes up to
$\chi^2$ terms even at the last step. The draw does not bound $R_j$ at earlier steps; on the trains we measured, where
$R_j$ is small on the last steps that carry the cost, the lossy cost then tracks the single-photon sampler under the same
loss. In the $x$-model, an emission $j$ in its own internal mode interferes with no other photon but
its label stays entangled with the train through the emitter, so it is not routed independently: its output row $r$ is
drawn from the exact marginal, jointly with the lost labels and the final emitter state, and the slot then carries the
fixed operator $G_j(r)=\sum_lU_{r,t_{j,l}}M^{(j)}_l$, a measured emission folded into the train like a lost one; the chain
runs on the remaining interfering emissions, and the photon is added at row $r$. The
correlations of, e.g., a state $\propto\ket{a_1b_2}+\ket{b_1a_2}$ with distinguishable photons 1 and 2 are kept, since
the second photon is drawn conditioned on the first.

\paragraph{Adaptive linear optics.} Uniform loss at the input, before the first adaptive detector, commutes with every
passive element and the adaptive detectors read what survives, so a lossy adaptive circuit is the lossless one on the
surviving photons; the policy sees the lossy counts. Loss inserted after an adaptive detector is not of this form: the
detector's earlier result and the policy it triggered cannot be reproduced by thinning the input. In
the $x$-model a shot partitions the surviving photons into interfering species and singletons. The species evolve
independently through the same unitaries $U^{p_{\le\lambda}}$, and the counts on the adaptive modes are sums over species:
stage~1 draws each species' level counts from its own marginal table (or its own forced chain), a singleton landing on an
adaptive mode with its renormalised single-photon probability, and the counts are added before the policy sees them;
stage~2 runs one forced-prefix chain per species. Given the partition the species are independent boson samplers on the
branch unitary, so the sampler is exact. Detector inefficiency on the adaptive detectors alone (a photon missed there and
nowhere else) is not covered: the policy would receive a thinned count while the chain is conditioned on the true one.

\paragraph{Non-linear sites.} Loss before the sites is the input-side draw above, and each shot then expands the response
for its own photon count, so the rejection factor $\|c\|_1^2$ is that of the surviving photons; loss after the last site is a
binomial thinning of the counts. Loss between two sites is not covered: a photon lost there has already shifted the phase
of the earlier sites, and tracing it couples branches through the Gram matrix of Sec.~\ref{sec:nlbs-fail}. For
distinguishability the site acts on the total photon number of its mode whatever the internal state, so
$e^{i\theta\hat n_x}=\prod_\mu e^{i\theta\hat n_{x,\mu}}$ over species: given the branch the species are independent and share
the branch index, the amplitude of a species-resolved output is $\sum_bc_b\prod_\mu\per U_b[s_\mu,t_\mu]/\sqrt{\prod_\mu s_\mu!}$,
and the Cauchy--Schwarz bound holds with $a_b=\prod_\mu\per U_b[s_\mu,t_\mu]$, so the acceptance stays $1/\|c\|_1^2$:
the extension of Algorithm~\ref{alg:nlbs} draws $b$ once, takes one Fock sample of $U_b$ per species, and accepts on the
species-resolved amplitudes.

Every case above is checked against an enumerated mixture over the loss and species partitions (Table~\ref{tab:valid} and
the tests named in the comments of the source).

\section{Leave-one-out values by division}
\label{sec:division}

The leave-one-out values of Sec.~3 of the main text, $e_{j-1}(a\setminus a_x)$, are the coefficients of
$F(\lambda)/(1+\lambda a_x)$ with $F(\lambda)=\prod_{i\in T_j}(1+\lambda a_i)$. Dividing is cheaper than the prefix--suffix
products of the main text but stable only in the direction that contracts: forward, $c_d=F_d-a_xc_{d-1}$, for
$|a_x|\le1$, and backward from the top degree, $c_{d-1}=(F_d-c_d)/a_x$, for $|a_x|>1$. Against the prefix--suffix
form, on Glynn vectors of $j$ Haar rows with $n=2j$ columns, forward division alone reaches relative errors of
$10^{-10}$ at $j=20$, $3\cdot10^{-5}$ at $j=40$ and $30$ at $j=50$; the mixed rule stays at $2\cdot10^{-13}$,
$5\cdot10^{-10}$ and $3\cdot10^{-8}$. Our GPU build uses it where the
division-free form is not instantiated.

\section{Formal proofs in Lean}
\label{sec:lean}

Lemma~1, Propositions~1 and~2 and stage~2 of the adaptive sampler (Sec.~5) of the main text are formalised in
Lean~4 with the Mathlib library. No step is left
unproved, and the proofs use only Lean's standard axioms (propositional extensionality, choice and quotient
soundness). Table~\ref{tab:lean} lists the correspondence; the Lean sources are archived at \href{https://doi.org/10.5281/zenodo.24023537}{doi:10.5281/zenodo.24023537}. Probabilities are real numbers, amplitudes complex numbers,
and slots, output modes and emission labels arbitrary finite types; nothing depends on floating-point arithmetic.

\begin{table}[H]\centering\footnotesize
\setlength{\tabcolsep}{4pt}\begin{tabular}{L{3.0cm}L{3.8cm}L{7.2cm}}\toprule
paper & Lean file & main declarations and content\\\midrule
Definition~1, Lemma~1 & \texttt{Basic.lean} & \texttt{Weights}: non-negative $f_j(z;T)$ with (C1), (C2).
\texttt{joint\_invariant}: $\Pr[z_{\le j},T_j=T]=\binom N{s_j}^{-1}\frac{s_0!}{s_j!}f_j(z_{\le j};T)$, the induction of
the proof. (a): \texttt{marginal\_sum\_eq\_one}, \texttt{marginal\_append}; (b): \texttt{output\_law}; (c):
\texttt{posterior\_from\_joint}\\
Proposition~1, Remark~2 & \texttt{SuperposedSources.lean} & \texttt{superposedWeights}: $|\perrect A[z,T]|^2$ is
chain-consistent for any offset when $A$ has orthonormal columns; \texttt{superposed\_target}: the chain started from
$n-k$ sources outputs $z$ with probability $\frac{(n-k)!}{n!}|\perrect A[z,[n]]|^2$; \texttt{fock\_target}: $k=n$, the
Clifford--Clifford law\\
Sec.~4, weights & \texttt{EmitterWeight.lean} & \texttt{emitterWeights}: $f_j(z;T)=\bra{v_z}\rho_T\ket{v_z}$ is
chain-consistent for any normalised $\psi$ over label strings (an MPS being one case); \texttt{emitter\_target}: the
output law of $\psi$\\
Proposition~2 & \texttt{EmitterKraus.lean},\newline \texttt{EmitterKrausRank.lean} & \texttt{gapwise\_kraus}: any Kraus
decomposition of every inner gap, and any vectors and covectors with the Gram matrices of the end gaps, leave $\rho_T$
unchanged; \texttt{gapwise\_kraus\_minimal}: decompositions with at most $\min(\text{original},\chi^2)$ operators per inner
gap and $\min(\text{original},\chi)$ per end gap exist (spectral theorem on the Choi and Gram matrices)\\
Sec.~5, stage~2 & \texttt{Adaptive.lean},\newline \texttt{Exchangeable.lean} & \texttt{adaptive\_continuation}: the chain
started after the prefix $a$ from the posterior of part (c) and aborted at the first adaptive row outputs the continuation
$y$ with probability $\mathbf 1[y\cap\mathcal A=\emptyset]\,q(a\,y)/q_r(a)$, for any chain-consistent weights;
\texttt{conditional\_eq}: for an exchangeable law, conditioning on the counts $p$ of the adaptive modes and on ``first
$r$ rows $a^p$, no later adaptive row'' agree, with the counting factor $n!/((n-r)!\,p!)$ (\texttt{sum\_sameCounts\_eq});
\texttt{fock\_exchangeable}: the Fock law is exchangeable\\\bottomrule
\end{tabular}
\caption{Correspondence between the main text and the Lean formalisation (namespace \texttt{SetChain}).}\label{tab:lean}
\end{table}

\noindent\emph{Modelling choices.} The chain of Lemma~1 is formalised as in its proof: a uniform initial $s_0$-set, then
one slot added uniformly among the unused ones at each step, which is the law of the sets $T_j$ of a uniformly random
order. A step whose normalisation vanishes gets probability zero, and the proof shows it is reached with probability zero.
Proposition~2 is stated for a train already grouped by $T$: the prefix vectors, then for each kept emission the operators
of the gap before it and its own operator, then the suffix covectors.

\noindent\emph{Not formalised.} The equivalence between the slot-by-slot draw and a random order $\sigma$; the weighted start;
the splitting of an arbitrary set $T$ into its gaps and the identification of the grouped form with the $\rho_T$ of the
weight of Sec.~4; the product of the per-gap counts that gives the bound on $R_j$; the physical derivations (the
superposed-source amplitudes and the emission model, whose outcomes enter as the formulas of the main text); for the
adaptive sampler, stage~1 (the outcome law and its generating function), the suffix-table draw of the prefix set
(Sec.~\ref{sec:columns}; the proof assumes an exact posterior draw) and the identification of $q_r(a^p)$ with the binomial
moment $M[p]$ in the acceptance probability $P(p)/M[p]$; the non-linear sampler, unequal sources, loss and partial
distinguishability, and all cost statements.